\documentclass[journal,draftclsnofoot,onecolumn,12pt]{IEEEtran}

\usepackage{amsmath} 
\usepackage{amsthm}
\usepackage{amsfonts}
\usepackage{amssymb}
\usepackage{algorithmic}
\usepackage{amssymb}
\usepackage{algorithm}

\usepackage{braket}
\usepackage{bbm}
\usepackage{bm}
\usepackage{booktabs}

\usepackage{CJK}
\usepackage{color}
\usepackage{cases}
\usepackage{cite}
\usepackage{caption}

\usepackage{epsf} 
\usepackage{epsfig}

\usepackage{graphicx}
\usepackage{gensymb}

\usepackage[colorlinks,citecolor=red, linkcolor=blue, anchorcolor=green]{hyperref}

\usepackage{indentfirst}

\usepackage{multirow}
\usepackage{makecell}

\usepackage[figureright]{rotating}

\usepackage{subfigure}
\usepackage{setspace}

\usepackage{textcomp}

\usepackage{url}

\allowdisplaybreaks

\begin{document}

\def\mba{{\mathbf{a}}}
\def\mbb{{\mathbf{b}}}
\def\mbc{{\mathbf{c}}}
\def\mbd{{\mathbf{d}}}
\def\mbe{{\mathbf{e}}}
\def\mbf{{\mathbf{f}}}
\def\mbg{{\mathbf{g}}}
\def\mbh{{\mathbf{h}}}
\def\mbi{{\mathbf{i}}}
\def\mbj{{\mathbf{j}}}
\def\mbk{{\mathbf{k}}}
\def\mbl{{\mathbf{l}}}
\def\mbm{{\mathbf{m}}}
\def\mbn{{\mathbf{n}}}
\def\mbo{{\mathbf{o}}}
\def\mbp{{\mathbf{p}}}
\def\mbq{{\mathbf{q}}}
\def\mbr{{\mathbf{r}}}
\def\mbs{{\mathbf{s}}}
\def\mbt{{\mathbf{t}}}
\def\mbu{{\mathbf{u}}}
\def\mbv{{\mathbf{v}}}
\def\mbw{{\mathbf{w}}}
\def\mbx{{\mathbf{x}}}
\def\mby{{\mathbf{y}}}
\def\mbz{{\mathbf{z}}}
\def\mb0{{\mathbf{0}}}
\def\mb1{{\mathbf{1}}}

\def\mbA{{\mathbf{A}}}
\def\mbB{{\mathbf{B}}}
\def\mbC{{\mathbf{C}}}
\def\mbD{{\mathbf{D}}}
\def\mbE{{\mathbf{E}}}
\def\mbF{{\mathbf{F}}}
\def\mbG{{\mathbf{G}}}
\def\mbH{{\mathbf{H}}}
\def\mbI{{\mathbf{I}}}
\def\mbJ{{\mathbf{J}}}
\def\mbK{{\mathbf{K}}}
\def\mbL{{\mathbf{L}}}
\def\mbM{{\mathbf{M}}}
\def\mbN{{\mathbf{N}}}
\def\mbO{{\mathbf{O}}}
\def\mbP{{\mathbf{P}}}
\def\mbQ{{\mathbf{Q}}}
\def\mbR{{\mathbf{R}}}
\def\mbS{{\mathbf{S}}}
\def\mbT{{\mathbf{T}}}
\def\mbU{{\mathbf{U}}}
\def\mbV{{\mathbf{V}}}
\def\mbW{{\mathbf{W}}}
\def\mbX{{\mathbf{X}}}
\def\mbY{{\mathbf{Y}}}
\def\mbZ{{\mathbf{Z}}}

\def\mcA{{\mathcal{A}}}
\def\mcB{{\mathcal{B}}}
\def\mcC{{\mathcal{C}}}
\def\mcD{{\mathcal{D}}}
\def\mcE{{\mathcal{E}}}
\def\mcF{{\mathcal{F}}}
\def\mcG{{\mathcal{G}}}
\def\mcH{{\mathcal{H}}}
\def\mcI{{\mathcal{I}}}
\def\mcJ{{\mathcal{J}}}
\def\mcK{{\mathcal{K}}}
\def\mcL{{\mathcal{L}}}
\def\mcM{{\mathcal{M}}}
\def\mcN{{\mathcal{N}}}
\def\mcO{{\mathcal{O}}}
\def\mcP{{\mathcal{P}}}
\def\mcQ{{\mathcal{Q}}}
\def\mcR{{\mathcal{R}}}
\def\mcS{{\mathcal{S}}}
\def\mcT{{\mathcal{T}}}
\def\mcU{{\mathcal{U}}}
\def\mcV{{\mathcal{V}}}
\def\mcW{{\mathcal{W}}}
\def\mcX{{\mathcal{X}}}
\def\mcY{{\mathcal{Y}}}
\def\ncalZ{{\mathcal{Z}}}

\def\mbbA{{\mathbb{A}}}
\def\mbbB{{\mathbb{B}}}
\def\mbbC{{\mathbb{C}}}
\def\mbbD{{\mathbb{D}}}
\def\mbbE{{\mathbb{E}}}
\def\mbbF{{\mathbb{F}}}
\def\mbbG{{\mathbb{G}}}
\def\mbbH{{\mathbb{H}}}
\def\mbbI{{\mathbb{I}}}
\def\mbbJ{{\mathbb{J}}}
\def\mbbK{{\mathbb{K}}}
\def\mbbL{{\mathbb{L}}}
\def\mbbM{{\mathbb{M}}}
\def\mbbN{{\mathbb{N}}}
\def\mbbO{{\mathbb{O}}}
\def\mbbP{{\mathbb{P}}}
\def\mbbQ{{\mathbb{Q}}}
\def\mbbR{{\mathbb{R}}}
\def\mbbS{{\mathbb{S}}}
\def\mbbT{{\mathbb{T}}}
\def\mbbU{{\mathbb{U}}}
\def\mbbV{{\mathbb{V}}}
\def\mbbW{{\mathbb{W}}}
\def\mbbX{{\mathbb{X}}}
\def\mbbY{{\mathbb{Y}}}
\def\nbbZ{{\mathbb{Z}}}

\def\mfrakR{{\mathfrak{R}}}

\def\mrma{{\rm a}}
\def\mrmb{{\rm b}}
\def\mrmc{{\rm c}}
\def\mrmd{{\rm d}}
\def\mrme{{\rm e}}
\def\mrmf{{\rm f}}
\def\mrmg{{\rm g}}
\def\mrmh{{\rm h}}
\def\mrmi{{\rm i}}
\def\mrmj{{\rm j}}
\def\mrmk{{\rm k}}
\def\mrml{{\rm l}}
\def\mrmm{{\rm m}}
\def\mrmn{{\rm n}}
\def\mrmo{{\rm o}}
\def\mrmp{{\rm p}}
\def\mrmq{{\rm q}}
\def\mrmr{{\rm r}}
\def\mrms{{\rm s}}
\def\mrmt{{\rm t}}
\def\mrmu{{\rm u}}
\def\mrmv{{\rm v}}
\def\mrmw{{\rm w}}
\def\mrmx{{\rm x}}
\def\mrmy{{\rm y}}
\def\mrmz{{\rm z}}

\newtheorem{lemma}{Lemma}
\newtheorem{definition}{Definition}
\newtheorem{remark}{Remark}
\newtheorem{theorem}{Theorem}
\newtheorem{proposition}{Proposition}
\newtheorem{corollary}{Corollary}
\newtheorem{example}{Example}
\newtheorem{assumption}{Assumption}

\newcommand{\upcite}[1]{\textsuperscript{\textsuperscript{\cite{#1}}}} 
\newcommand{\Romann}[1]{\uppercase\expandafter{\romannumeral #1}} 
\newcommand{\romann}[1]{\expandafter{\romannumeral #1}} 
\newcommand{\colorb}[1]{{\color{blue} #1}} 
\newcommand{\colorr}[1]{{\color{red} #1}}


\newcommand{\ceil}[1]{\lceil #1\rceil} 

\def\argmin{\operatorname{arg~min}}
\def\argmax{\operatorname{arg~max}}

\def\sinc{{\rm sinc}}
\def\cosc{{\rm cosc}}

\def\larrow{\leftarrow}
\def\rarrow{\rightarrow}
\def\triequ{\triangleq}
\def\simequ{\simeq}

\def\figref#1{Fig.\,\ref{#1}}%
\def\tabref#1{Table\,\ref{#1}}%
\def\equref#1{(\ref{#1})}%
\def\appref#1{Appendix\,\ref{#1}}%
\def\lemref#1{Lemma\,\ref{#1}}%
\def\defref#1{Definition\,\ref{#1}}%
\def\theref#1{Theorem\,\ref{#1}}%
\def\remref#1{Remark\,\ref{#1}}%
\def\secref#1{Sec.\,\ref{#1}}%

\def\assref#1{Assumption\,\ref{#1}}%
\def\ie{{\em i.e.}}
\def\eg{{\em e.g.}}
\def\rme{{\rm e}}
\def\rmd{{\rm d}}
\def\dB{{\rm dB}}
\def\x{\times}
\def\T{\intercal}
\def\H{\dagger}
\def\wbar{\overline}
\def\what{\widehat}
\def\d{{\rm d}}
\def\E{{\mathbb E}}
\def\pd{\partial}
\def\e{{\rm e}}
\def\1{\mathbbmtt{1}}
\def\var{\operatorname{Var}}
\def\cov{\operatorname{Cov}}
\def\mean{\operatorname{mean}}
\def\P{{\mathbb P}}

\def\R{{\mathbb R}}

\def\erfc{\operatorname{erfc}}
\def\erf{\operatorname{erf}}
\def\opt{\mathrm{opt}}

\def\sinr{\mathtt{SINR}}   
\def\snr{\mathtt{SNR}}
\def\sir{\mathtt{SIR}}
\def\scnr{\mathtt{SCNR}}
\def\ase{\mathtt{ASE}}
\def\se{\mathtt{SE}}

\title{State-of-the-Art Underwater Vehicles and Technologies Enabling Smart Ocean: Survey and Classifications}

\author{
    \IEEEauthorblockN{
                      Jiajie Xu$^{1}$, ~\IEEEmembership{Member,~IEEE},
                      Xabier Irigoien$^{2}$,
                      Mohamed-Slim Alouini$^{1}$, ~\IEEEmembership{Fellow,~IEEE}
    }
    \thanks{
            \IEEEauthorblockA{$^{1}$ Electrical and Computer Engineering (ECE), Computer, Electrical, and Mathematical Science and Engineering (CEMSE) Division, King Abdullah University of Science and Technology (KAUST), Thuwal, 23955-6900, Kingdom of Saudi Arabia}. 
             \IEEEauthorblockA{$^{2}$  OSSARI, NEOM, Kingdom of Saudi Arabia}.
            (e-mail: jiajie.xu.1@kaust.edu.sa, mustafa.kishk@mu.ie, xabier.irigoyen@neom.com, slim.alouini@kaust.edu.sa.) This work was supported by the ERIF/OSSARI Funding.
    }
}

\maketitle

\begin{abstract}

The exploration and sustainable use of marine environments have become increasingly critical as oceans cover over 70\% of the Earth's surface. This paper provides a comprehensive survey and classification of state-of-the-art underwater vehicles (UVs) and supporting technologies essential for enabling a smart ocean. We categorize UVs into several types, including remotely operated vehicles (ROVs), autonomous underwater vehicles (AUVs), hybrid underwater vehicles (HUVs), unmanned surface vehicles (USVs), and underwater bionic vehicles (UBVs). These technologies are fundamental in a wide range of applications, such as environmental monitoring, deep-sea exploration, defense, and underwater infrastructure inspection. Additionally, the paper explores advancements in underwater communication technologies—namely acoustic, optical, and hybrid systems—as well as key support facilities, including submerged buoys, underwater docking stations, and wearable underwater localization systems. By classifying the vehicles and analyzing their technological capabilities and limitations, this work aims to guide future developments in underwater exploration and monitoring, addressing challenges such as energy efficiency, communication limitations, and environmental adaptability. The paper concludes by discussing the integration of artificial intelligence and machine learning in enhancing the autonomy and operational efficiency of these systems, paving the way for the realization of a fully interconnected and sustainable Smart Ocean.

\end{abstract}


\section{Introduction}
The oceans, covering more than 70\% of Earth's surface, play a critical role in regulating climate, supporting biodiversity, and enabling global trade. However, the vast depths and harsh underwater environments present significant challenges for exploration, monitoring, and sustainable use of marine resources. Over the past few decades, advancements in underwater vehicle technologies have revolutionized our ability to explore and manage these hidden realms. 


Underwater exploration can enrich the critical resources for human beings extremely in terms of food, economy, fuel, etc. Compared with terrestrial activities, which are assisted with various developed tools, vehicles, and methods, underwater activities are underdeveloped, and the tools available to them are scarce. In the past decades, different types of underwater vehicles (UVs) and technologies have been developed by both industrial companies and research institutes. Various UVs are applied for underwater data collection, monitoring, communication, remote operation, etc. According to the purpose and abilities, UAs can be clarified as (i) Remotely operated vehicles (ROVs); (ii) Autonomous underwater vehicles (AUVs); (iii) Hybrid underwater vehicles (HUVs), including hybrid ROV-AUVs and hybrid UAV\footnote {Unmanned aerial vehicle}-AUVs; (iv) Unmanned surface vehicles (USVs); (v) Underwater gliders (UGs); (vi) Underwater bionic vehicles (UBVs); (vii) Underwater special robots (USRs).
Besides UVs, promising underwater technologies include underwater wireless communications, underwater sensing systems, and various supporting facilities.


There are some surveys available on underwater wireless acoustic communications (UWACs) \cite{aliRecentAdvancesFuture2020,stojanovic2006underwater}, underwater wireless optical communications (UWOCs) \cite{schirripaspagnoloUnderwaterOpticalWireless2020a,7450595, zengSurveyUnderwaterOptical2016}, underwater wireless sensor networks (UWSNs) 
\cite {ghafoorOverviewNextGenerationUnderwater20191,gkikopouli2012survey, han2012localization}, underwater internet of things (UIoTs) \cite{qiuUnderwaterInternetThings2020a, bello2022internet ,mohsan2022towards}. However, few papers focus on the development of UVs and the promising abilities they provide for underwater explorations.

This paper explores the classifications, current technological advancements, and critical design challenges associated with underwater vehicles as well as some support facilities in the context of enabling a smart ocean. It examines how these technologies contribute to diverse applications such as deep-sea exploration, environmental monitoring, underwater infrastructure inspection, and defense. Moreover, the analysis highlights the growing integration of artificial intelligence (AI), machine learning, and advanced sensing techniques that enhance vehicle autonomy and operational efficiency. However, several design challenges, including energy efficiency, communication limitations in deep water, and the robustness of vehicle components to withstand harsh marine conditions, remain at the forefront of ongoing research.

By classifying these vehicles and analyzing their technological capabilities and limitations, this paper aims to provide a comprehensive overview of the current state of underwater vehicles. Furthermore, it addresses the design challenges that must be overcome to fully realize the vision of a Smart Ocean, where real-time monitoring, sustainable resource management, and secure operations can be achieved through seamless underwater technological innovations. 

Before the survey and the classifications of underwater vehicles, a widely investigate of the companies in the underwater field and their products are presented in \tabref{table_companies}.


\section{State-of-the-Art Maritime/Underwater Vehicles}

\begin{figure*}[htbp]  
\centering
    \includegraphics[width=15cm]{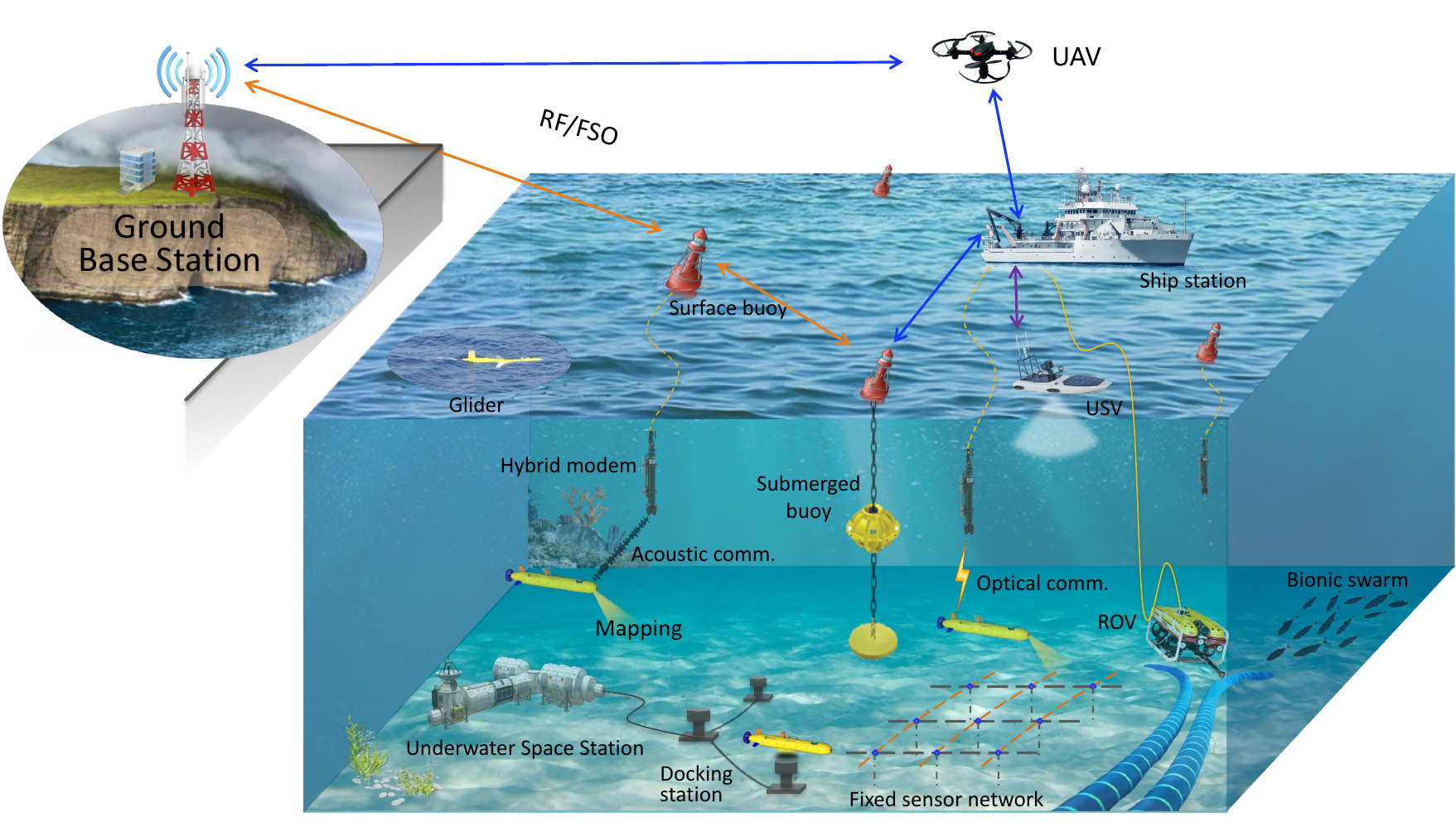}
    \caption{Description of the future smart ocean}
    \label{fig_system_model}
\end{figure*}

A typical description of the future smart ocean is shown in \figref{fig_system_model} where the hardware systems are emphasized.
UAVs, surface buoys, and ships bridge the connection between the terrestrial and underwater worlds. USVs, gliders, ROVs, AUVs, bionic swarms, underwater docking, and sensor networks are deployed to execute underwater exploration. It is easy to see that since human beings are not the masters of underwater activities, UVs play a critical role in the applications. This section provides details of the current applied maritime/underwater vehicles.

\begin{sidewaystable}[h]\small
\tabcolsep=1.2mm
\centering
\caption{Underwater devices and related companies}\vspace{0mm}
	\begin{tabular}{c|c|c|c|c|c|c|c|c|c|c|c|c|c|c}  
	    \toprule
	     Company & Country &  ROV & AUV & USV & Glider & \makecell{Hybrid\\AUV} & \makecell{Bio-\\Robots} & \makecell{UW\\Swarm} & \makecell{Optical\\Modem} & \makecell{Acoustic\\Modem} & \makecell{Docking\\System} & \makecell{UW\\Camera} & \makecell{Navigation\\Localization} & \makecell{Mapping\\Detection}\\
	    \midrule
	    \makecell{Kongsberg\\Maritime \cite{Kongsberg_Maritime}}  & Norway& & \checkmark  &  \checkmark  &  &   &    &  &   &  \checkmark  &  &   &  \checkmark  & \checkmark \\ \hline
	    Teledyne Marine \cite{teledynemarine}   & USA &\checkmark & \checkmark  &  \checkmark  & \checkmark &   &    &  &   &  \checkmark  &  & \checkmark  &  \checkmark  &  \\ \hline
	    SUBNORE \cite{subnero}  & Singapore   & &   &    &  &   &    &  &   &  \checkmark  &  &   &    &  \\ \hline
        Popoto Modem \cite{popotomodem}  & USA  & &   &    &  &   &    &  &   &  \checkmark  &  &   &    &  \\ \hline
        Hydromea \cite{hydromea}  & Switzerland & \checkmark &   &    &  &   &    & \checkmark & \checkmark  &    &  &   &    &  \\ \hline
        Nortek Group \cite{nortekgroup}  & Norway &  &   &    &  &   &    &  &   &  \checkmark  &  & \checkmark  &  \checkmark  & \checkmark \\ \hline
        Sonardyne \cite{sonardyne}  & Uk &  &   &    &  &   &    &  & \checkmark  &  \checkmark  &  &  &  \checkmark  & \checkmark \\ \hline
        Evologics \cite{evologics}   & Germany & &   &  \checkmark  &  &   &  \checkmark  &  &   &  \checkmark  &  &   &  \checkmark  & \checkmark \\ \hline
        Subsea Tech \cite{subseatech}  & France &\checkmark & \checkmark  & \checkmark  &  &   &    &  &   &    & $-$ & \checkmark  &    &  \\ \hline
        Shimadzu \cite{shimadzu}  & Japan  & &  &    &  &   &    &  & \checkmark  &    &  &   &    &  \\ \hline
        ACTEON \cite{acteon}  & UK & \checkmark &  &    &  &   &   &  &   &    & &  &    & \checkmark \\ \hline
        ECA Group \cite{ecagroup}  & France &\checkmark & \checkmark  &  \checkmark  & &   &    &  &   &  \checkmark  &  &  &  \checkmark  & \checkmark \\ \hline
        OceanEco \cite{oceaneco}   & China  &\checkmark & \checkmark &    &  &   &    &  &   &    &  &   &    &  \\ \hline
        Pengpai Ocean \cite{pengpaix}  & China  &\checkmark & \checkmark &    &  &   &    &  &   &    &  &   &    &  \\ \hline
        RoboSea \cite{robosea}  & China &\checkmark & \checkmark &    &  &   &  \checkmark  &  &   &    &  &   &    &  \\ \hline
        \makecell{T-SEA Marine\\Technology \cite{t-sea}}  & China  &\checkmark & \checkmark &    & \checkmark &   &    &  &   & \checkmark   &  & \checkmark  &  \checkmark  & \checkmark \\ \hline
        SJTU-UEI \cite{sjtu} & China  &\checkmark & \checkmark &    & \checkmark &   &    &  &   &    &  & \checkmark  &    &  \\ \hline
        
        GeoSpectrum \cite{geospectrum}  & Canada & &  &    &  &   &    &  &   &  \checkmark  &  &   &  \checkmark  &  \\ \hline
        THALES Group \cite{thalesgroup}  & France  & &  &  \checkmark  &  &   &    &  &   &  \checkmark  &  &   &   \checkmark &  \\ \hline
        L3 Harris \cite{l3harris}  & USA & & \checkmark &  \checkmark  &  &   &    &  &   &    &  &   &  \checkmark  &  \\ \hline
        Elbit Systems \cite{elbitsystems}  & USA & & \checkmark &  \checkmark  &  &   &    &  &   &  \checkmark  &  &   &  \checkmark  &  \checkmark\\ \hline
        Textron Systems \cite{textronsystems}  & USA & &  &  \checkmark  &  &   &    &  &   &    &  &   &    &  \\ \hline
        Atlas Elektronik \cite{atlaselektronik}  &  UK &&  &  \checkmark  &  &   &    &  &   &  \checkmark  &  &   &    &  \checkmark\\ \hline
        \makecell{Smart Ocean\\Technology \cite{smartoceantech}}   &China  &  &  &    &  &   &    &  &   & \checkmark &  &   &  \checkmark  &  \\ 
    	\bottomrule 
	\end{tabular}	
\label{table_companies}
\leftline{ UW is the abbreviation of underwater, \checkmark \ means the product is available, $-$ means the company is working on the product.}
\end{sidewaystable}


\clearpage

As the most popular UVs that are applied to help underwater exploration, ROVs, and AUVs have developed a lot with the attention and investment of various commercial companies. Here, we classify ROVs and AUVs according to their capabilities and summarize them as shown in \tabref{111}.

\begin{table}[h]
    \centering
    \begin{tabular}{cccccc}
    \toprule
     Name &  Size & Weight & Payload & Depth & Mission duration \\
      \midrule
      \multirow{3}*{ROV }  & Large  & $>$1000 kg & $>$ 100 kg &  $>$ 2000 m & - \\ \cline{2-6}
                         ~ & Medium & $\leq$ 1000 kg & $\leq$ 100 kg &  $\leq$ 2000 m & - \\\cline{2-6}
                          ~ & Small & $<$ 50 kg & $<$ 10 kg &  $<$ 200 m & - \\ \hline
       \multirow{3}*{AUV }  & Large &  $>$ 500 kg   & $>$ 10 kg & $>$ 500 m &  Days \\ \cline{2-6}
                          ~ & Medium & $\leq$ 500 kg  & $\leq$ 10 kg & $\leq$ 500 m& Decades Hours \\ \cline{2-6}
                        ~ &  Small   & $<$ 100 kg  & $<$ 3 kg & $<$ 50 m& Hours \\    
    \bottomrule
    \end{tabular}
    \caption{Classification of ROVs and AUVs}
    \label{111}
\end{table}

\subsection{Large / Medium / Small ROVs}

ROVs are tethered to a surface vessel or platform via a cable, allowing human operators to control them in real time. The tether provides power and communication, enabling the operator to navigate the vehicle, manipulate objects, and receive immediate feedback through video and sensor data. ROVs are ideal for tasks requiring precise control and the ability to respond instantly to environmental conditions.

\begin{figure}[ht]	
\centering
	\subfigure[ROV hired by Global Foundation for Ocean Exploration \cite{rov111}]{
	    \begin{minipage}[b]{0.48\textwidth}
	    \includegraphics[width=1\textwidth]{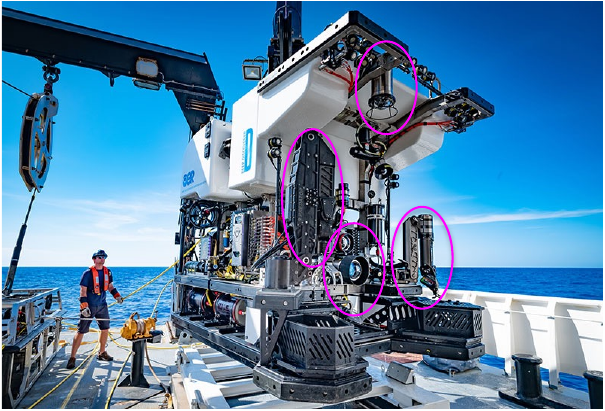}
		\end{minipage}
		\label{fig_large_rov_1}}
    \subfigure[Chinese Jiao Long \cite{jiaolong}]{
    	\begin{minipage}[b]{0.399\textwidth}
   		\includegraphics[width=1\textwidth]{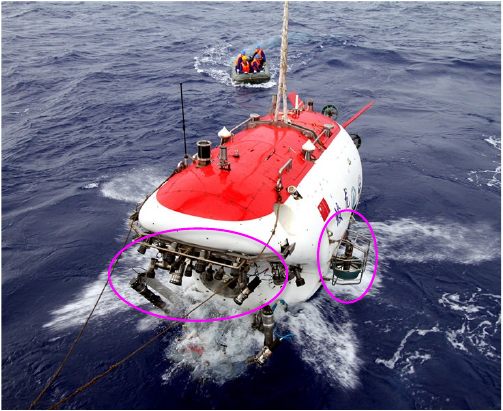}
    	\end{minipage}
		\label{fig_Chinese_Jiao_Long}}
\caption{Large ROVs}
\label{fig_large_rov}
\end{figure}

Two large ROVs are shown in \figref{fig_large_rov}, in which one is hired by Global Foundation for Ocean Exploration \cite{rov111}, one is developed by China and used for deep sea exploration \cite{jiaolong}, Jiao Long.
The large ROV can process hard work underwater, normally in the deep sea. 
A large ROV normally is costly. However, it has a large payload capacity, and it is resistant to water flow impact, which can work in unstable underwater environments. 

The large ROV can, for example, (i) do underwater mining, (ii) do pipeline laying, and (iii) do other engineering operations. Since they usually work in the deep sea, pressure resistance, resistance to water flow impact, and the stability of the entire robot-controlled system become key points in the design. The Jiao Long ROV can take samples from Mariana Trench at a depth of around 11000 meters, which makes it the first full-sea-depth working ROV. However, from a cost perspective, the working life of deep-sea ROVs is limited, and their pressure-resistant components will suffer varying degrees of wear during each deep dive, so the cost of using deep-sea robots is high.

\begin{figure}[ht]	
\centering
	\subfigure[The medium ROV working for simple operation \cite{rov_defender}]{
	    \begin{minipage}[b]{0.45\textwidth}
	    \includegraphics[width=1\textwidth]{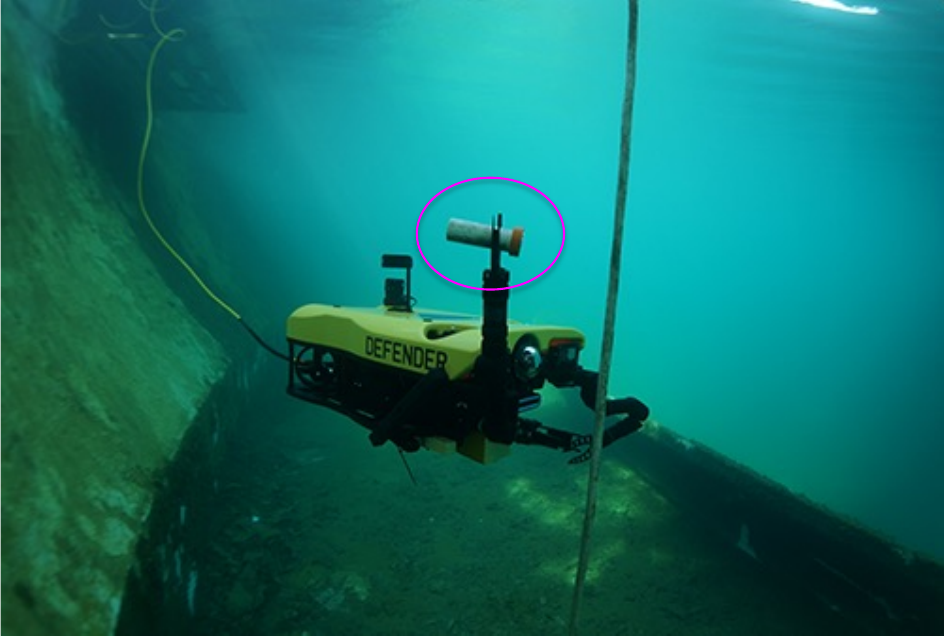}
		\end{minipage}
		\label{fig_medium rov 1}}
    \subfigure[The medium ROV working for monitoring and simple operation \cite{argus_rs}]{
    	\begin{minipage}[b]{0.391\textwidth}
   		\includegraphics[width=1\textwidth]{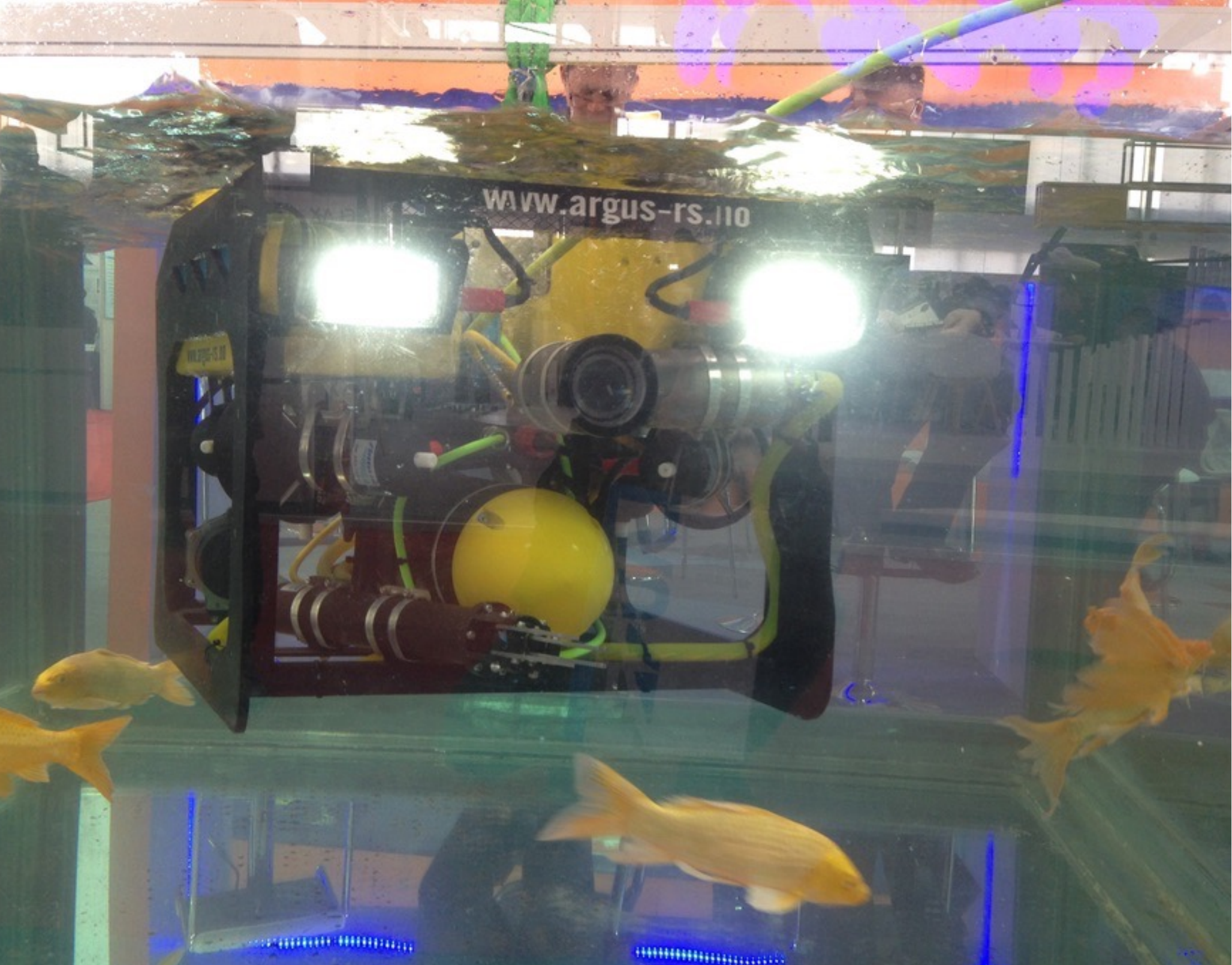}
    	\end{minipage}
		\label{fig_medium rov 2}}
\caption{Medium ROVs}
\label{fig_medium_rov}
\end{figure}

\begin{figure}[ht]	
\centering
	\subfigure[Open source small ROV developed by Blue Robotics \cite{bluerov2}]{
	    \begin{minipage}[b]{0.45\textwidth}
	    \includegraphics[width=1\textwidth]{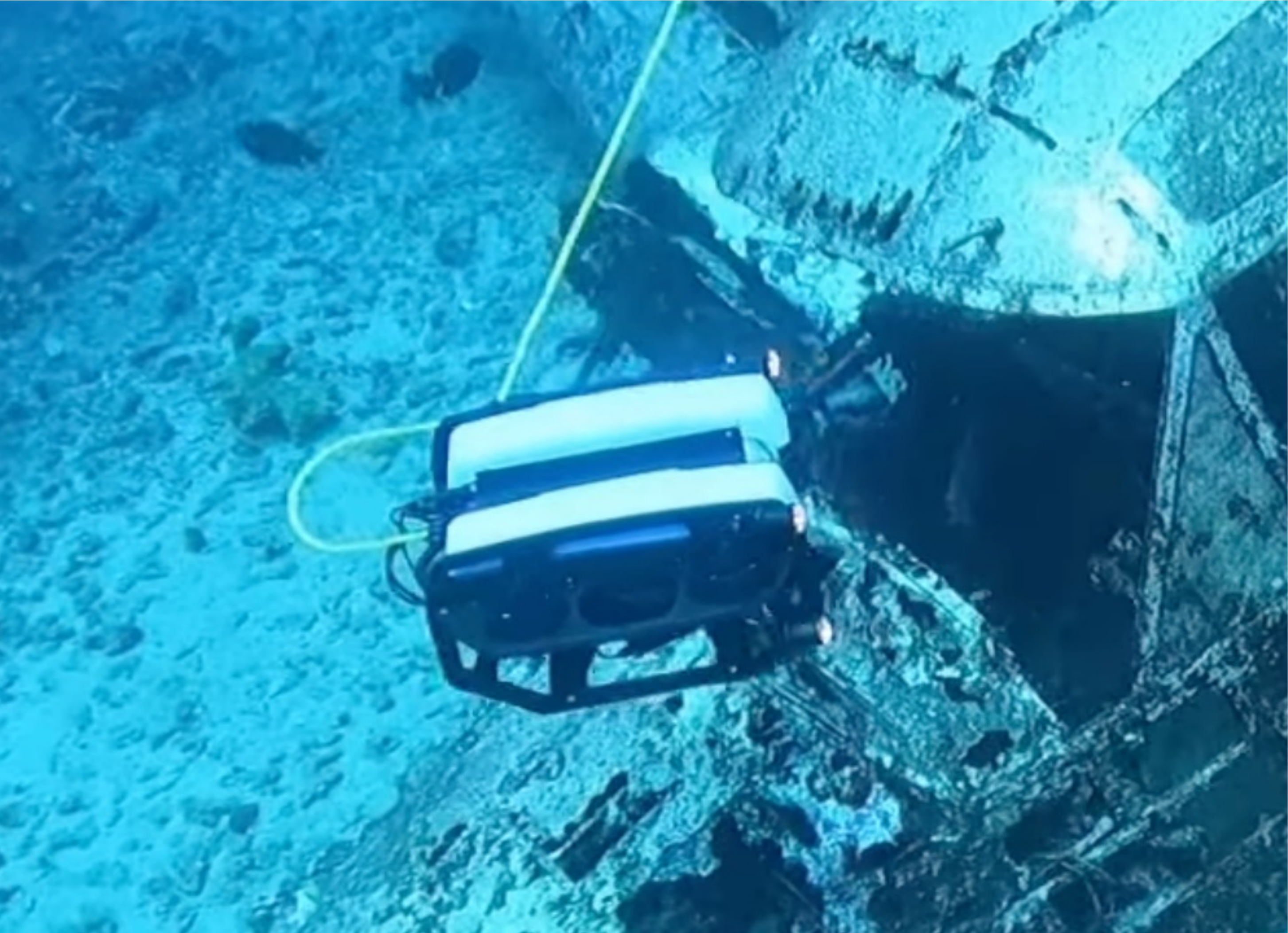}
		\end{minipage}
		\label{fig_small_rov_1}}
    \subfigure[Small ROV developed by QYSEA China \cite{fifish}]{
    	\begin{minipage}[b]{0.427\textwidth}
   		\includegraphics[width=1\textwidth]{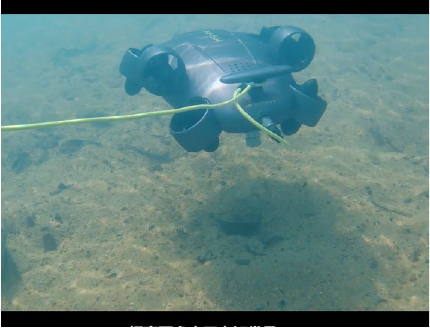}
    	\end{minipage}
		\label{fig_small_rov_2}}
\caption{Small ROVs}
\label{fig_small_rov}
\end{figure}

Medium and small ROVs are normally equipped with underwater cameras, underwater lights, and robotic arms, which can be applied for underwater videos, underwater monitoring, or sample underwater operation, for example, (i) underwater pipeline inspections, (ii) underwater infrastructure damage detection; (iii) dam crack detection; (iv) waterway silt or blockage detection; (v) underwater auxiliary engineering operations. Templates of medium and small ROVs are shown in \figref{fig_medium_rov} and \figref{fig_small_rov}.

\subsection{Large / Medium / Small AUVs}

In contrast to ROVs, AUVs operate independently without a physical connection to the surface. They are pre-programmed with specific missions and can make real-time decisions based on onboard sensor inputs. AUVs are well-suited for systematic survey mapping and repetitive data collection over large areas, as they can navigate to greater depths and cover more extensive distances without direct human intervention. The advantages of AUVs can be summarized as: (i) unmanned; (ii) no cables, no tangling; (iii) no limitation of activity scope; (iv) no required surface assistance vessel; (v) hard to be detected in underwater; 

\begin{figure}	
\centering
	\subfigure[The large AUV applied by Chinese Goverment]{
	    \begin{minipage}[b]{0.476\textwidth}
	    \includegraphics[width=1\textwidth]{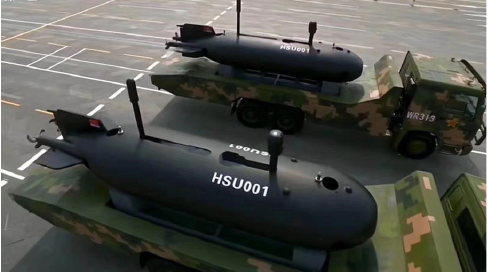}
		\end{minipage}
		\label{fig_large_auv_1}}
    \subfigure[The large AUV applied by NOAA \cite{auv_noaa}]{
    	\begin{minipage}[b]{0.4\textwidth}
   		\includegraphics[width=1\textwidth]{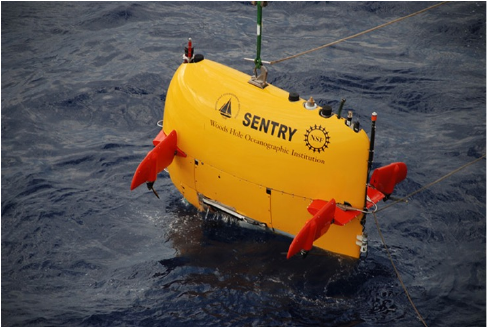}
    	\end{minipage}
		\label{fig_large_auv_2}}
\caption{Large AUVs}
\label{fig_large_auv}
\end{figure}

\begin{figure*}[htbp]  
\centering
    \includegraphics[width=13cm]{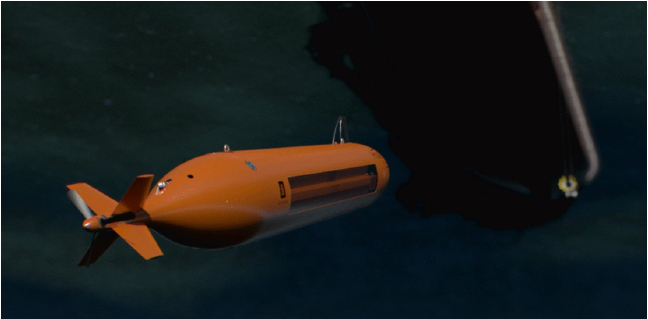}
    \caption{The large AUV in the shape of torpedo \cite{auv_kongsberg}}
    \label{fig_auv_torpedo}
\end{figure*}

Three practical large AUVs in different shapes are shown in \figref{fig_large_auv} and \figref{fig_auv_torpedo}. Considering the high cost and the strict requirements for system stability, large AUVs are usually used in the military field, such as terrain surveys, enemy submarine reconnaissance, etc. This has also directly led to the fact that there are few countries in the world that can develop and use such large AUVs.

Compared with large AUVs, medium and small AUVs are more useful in the field of civilian applications. 

\begin{figure}	
\centering
	\subfigure[Medium AUV in the shape of torpedo developed by WHOI \cite{auv_whoi}]{
	    \begin{minipage}[b]{0.5\textwidth}
	    \includegraphics[width=1\textwidth]{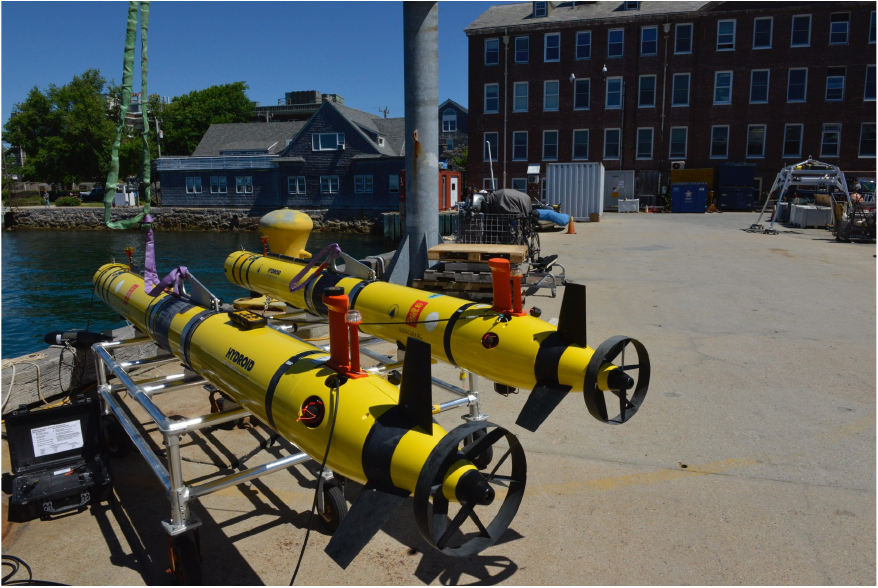}
		\end{minipage}
		\label{fig_medium_auv}}
    \subfigure[Small AUV in the shape of torpedo developed by Xi'an Tianhe \cite{auv_thtw}]{
    	\begin{minipage}[b]{0.367\textwidth}
   		\includegraphics[width=1\textwidth]{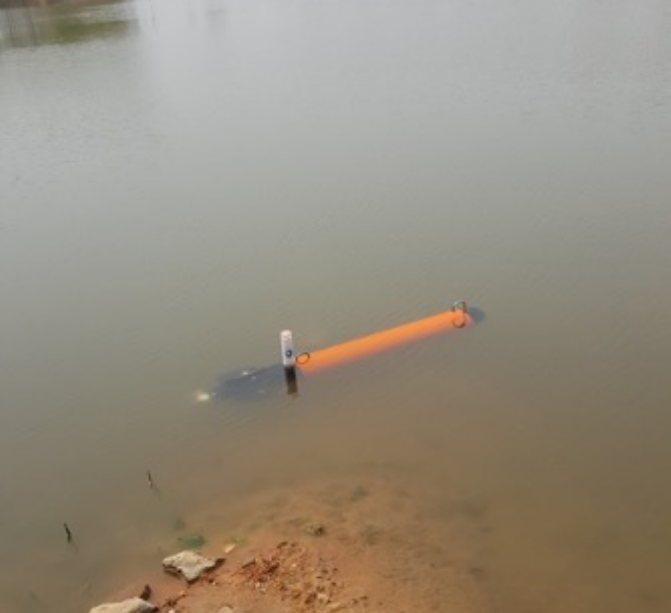}
    	\end{minipage}
		\label{fig_small_auv}}
\caption{Medium and small AUVs}
\label{fig_medium_small_auv}
\end{figure}

Medium and small AUVs are versatile tools in marine operations (as shown in \figref{fig_medium_small_auv}), offering a range of applications across scientific research, commercial endeavors, and military activities. Their compact size, maneuverability, and ability to operate in confined or shallow waters make them particularly valuable in various scenarios:
\begin{itemize}
    \item Hydrographic and Geophysical Surveys: These AUVs are instrumental in mapping the seafloor, conducting bathymetric surveys, and collecting geophysical data. Equipped with sensors like multibeam echosounders and side-scan sonars, they efficiently cover large areas, providing high-resolution data essential for nautical charting, coastal engineering, and offshore construction projects.
    \item Environmental Monitoring: Small and medium AUVs play a crucial role in monitoring water quality, marine biodiversity, and environmental changes. They can be deployed to collect samples, measure parameters such as temperature, salinity, and dissolved oxygen levels, and detect pollutants, contributing valuable data for climate studies, pollution assessment, and marine conservation efforts.
    \item Search and Rescue Operations: In search and rescue scenarios, these AUVs can be rapidly deployed to locate missing persons or objects underwater. Their ability to navigate challenging environments and provide real-time data enhances the efficiency and safety of rescue missions.
    \item Mine Countermeasures: Medium-sized AUVs are utilized in military applications to detect and neutralize underwater mines. Their deployment reduces risks to human divers and increases the effectiveness of mine-clearing operations.
    \item Inspection and Maintenance: These AUVs are employed to inspect underwater structures such as pipelines, cables, and offshore platforms. They can carry out detailed inspections, identify potential issues, and perform maintenance tasks, ensuring the integrity and longevity of marine infrastructure.
    \item Scientific Research: In polar regions, small AUVs have been deployed to study ice dynamics, collect data on sea ice thickness, and map under-ice topography, providing insights into climate change impacts.
\end{itemize}

As the two most indispensable tools for underwater exploration, the differences between ROV and AUV can be summarized as follows:
\begin{itemize}
    \item Control and Autonomy: ROVs require continuous human control via a tether, offering immediate responsiveness, whereas AUVs are designed for autonomous operation, making decisions based on pre-programmed instructions and sensor data.
    \item Mobility and Range: The tether limits ROVs' range and mobility, making them more suitable for localized tasks. AUVs, being untethered, can traverse larger areas and are ideal for extensive surveys and missions requiring broader coverage.
    \item Applications: ROVs are commonly employed in scenarios where precise manipulation or real-time data is crucial, such as in offshore energy operations or detailed marine research. AUVs are frequently used for tasks like oceanographic mapping, environmental monitoring, and long-duration missions where autonomy is advantageous.
    \item Understanding these differences is essential for selecting the appropriate vehicle type based on the specific requirements and objectives of the underwater mission.
\end{itemize}

\subsection{Hybrid Underwater Vehicles}

\begin{figure}	
\centering
	\subfigure[Hybrid AUV with a stern thruster]{
	    \begin{minipage}[b]{0.45\textwidth}
	    \includegraphics[width=1\textwidth]{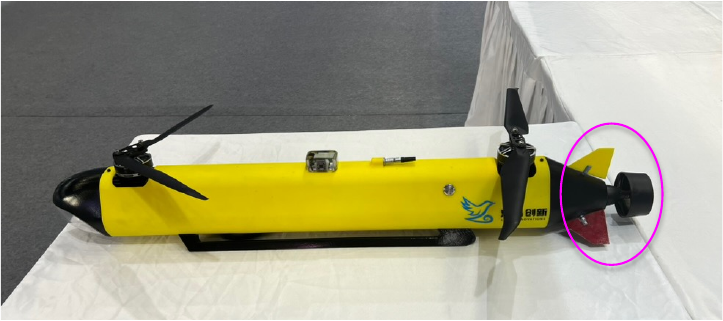}
		\end{minipage}
		\label{fig_hybrid auv 1}}
    \subfigure[Hybrid AUV with multiple thrusters]{
    	\begin{minipage}[b]{0.363\textwidth}
   		\includegraphics[width=1\textwidth]{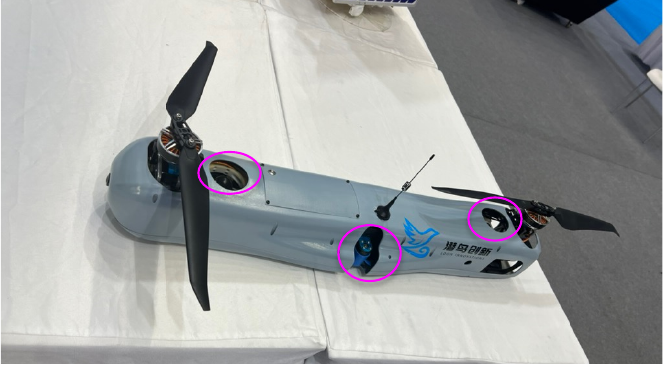}
    	\end{minipage}
		\label{fig_hybrid auv 2}}
\caption{Hybrid AUVs in the shape of torpedo}
\label{fig_hybrid auv}
\end{figure}

\begin{figure*}[htbp]  
\centering
    \includegraphics[width=9cm]{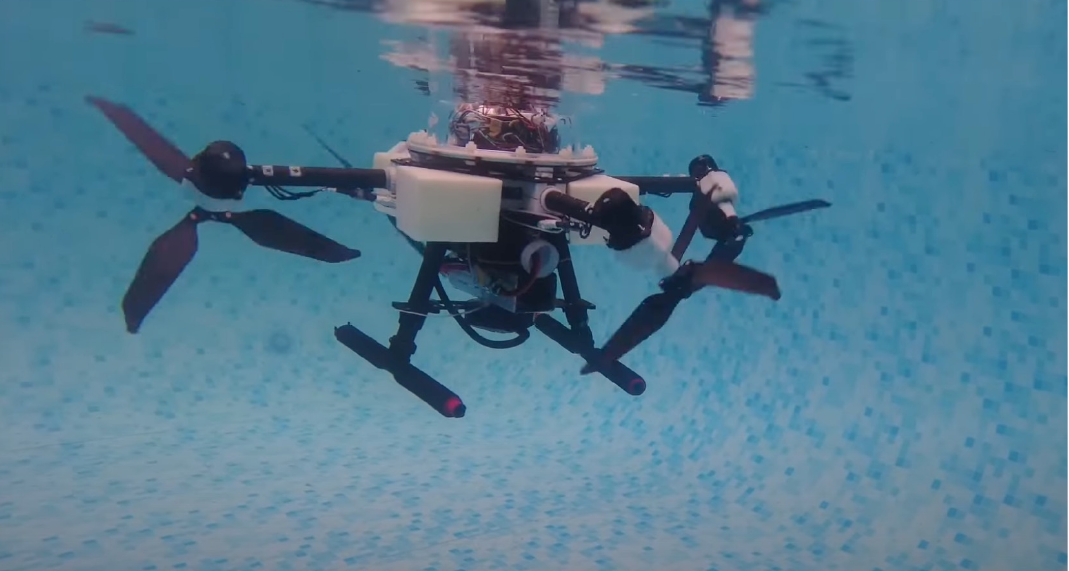}
    \caption{The hybrid AUV in the shape of quadcopter \cite{tongjiflyingfish}}
    \label{fig_hybrid auv 4}
\end{figure*}

Considering the underwater flexibility of AUVs and the disability above the water, hybrid AUVs (HAUVs) are developed that can realize cross-media work above and below water. As we know, there are no commercial products of HAUVs. There are only some experiments and demos that have been developed by some companies and universities. HAUVS can be designed in different shapes, for example, the shape of the torpedo and the shape of the quadcopter. 
The UAV-AUV, as one of the typical Hybrid UVs, can work as an AUV underwater and as a UAV in the air.
"Loon Innovations," a start-up company in China, developed two different kinds of HAUVs as shown in \figref{fig_hybrid auv}. The HAUV with a stern thruster (as shown in \figref{fig_hybrid auv 1}) is aimed at high moving speed ahead. The HAUV with multiple thrusters (as shown in \figref{fig_hybrid auv 2}) can achieve movements such as going up/down, left/right, and even turning in place.
"TJ-FlyingFish" is developed by Tongji University, China \cite{tongjiflyingfish,tongjiflyingfish2} (as shown in \figref{fig_hybrid auv 4}). "Nezha III (VTOL)" is developed by Shanghai Jiaotong University \cite{nezha1,nezha2}.

The ROV-AUV, one of the hybrid UVs, integrates the capabilities of ROVs and AUVs. The main goal is to create a more versatile, efficient, and capable vehicle that can adapt to different mission requirements. 
ROV-AUVs combine the strengths of AUVs and ROVs, enabling them to handle a wide range of underwater tasks—from long-duration, large-scale surveys to detailed manipulation and inspection tasks. While traditional AUVs are limited by battery life, ROV-AUVs can switch to tethered mode to extend their operation time when necessary. In tethered mode, ROV-AUVs can transmit high-bandwidth data in real-time, which is crucial for complex or critical tasks such as pipeline inspections or subsea construction. In ROV mode, operators can perform complex manipulations or inspections in real-time, providing higher accuracy in tasks requiring human oversight. By offering both autonomous and tethered operations, ROV-AUVs can reduce downtime associated with switching between different vehicles for different tasks, increasing overall operational efficiency.
ROV-AUVs can handle diverse underwater conditions, such as strong currents or varying depths, making them well-suited for a variety of missions in unpredictable or harsh environments.

\subsection{Underwater gliders}

\begin{figure*}[htbp]  
\centering
    \includegraphics[width=11cm]{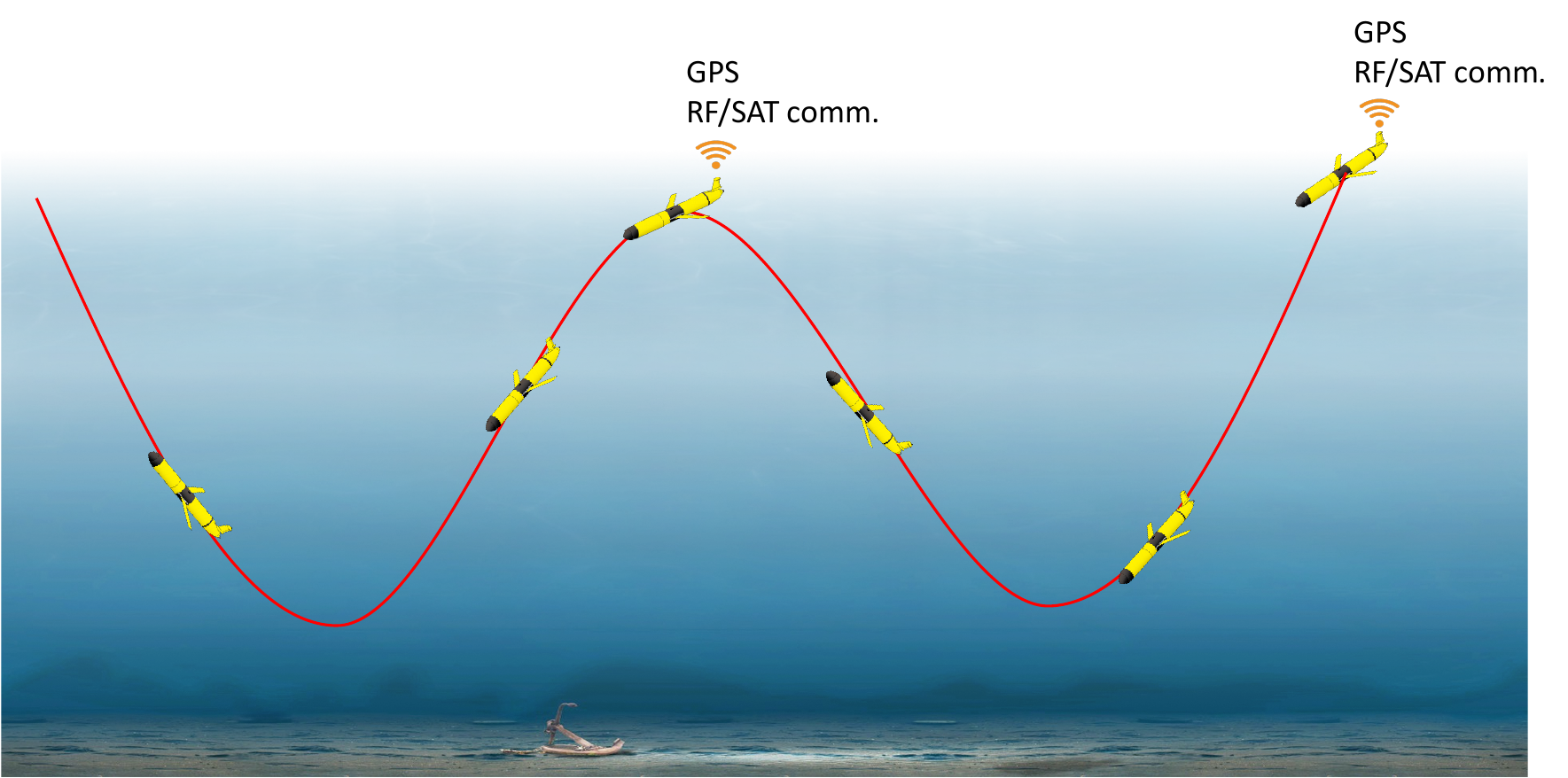}
    \caption{A diagram of the underwater glider movement mode}
    \label{fig_underwater_glider_3}
\end{figure*}

An underwater glider is a type of autonomous underwater vehicle (AUV) that moves through the water by changing its buoyancy, allowing it to "glide" through the ocean in a sawtooth pattern. This unique propulsion method makes underwater gliders highly energy-efficient, enabling long-term ocean observations over vast distances. A diagram of the underwater glider movement mode is shown in \figref{fig_underwater_glider_3}.

\begin{figure}	
\centering
	\subfigure[Underwater glider from WHOI \cite{whoi_glider}]{
	    \begin{minipage}[b]{0.467\textwidth}
	    \includegraphics[width=1\textwidth]{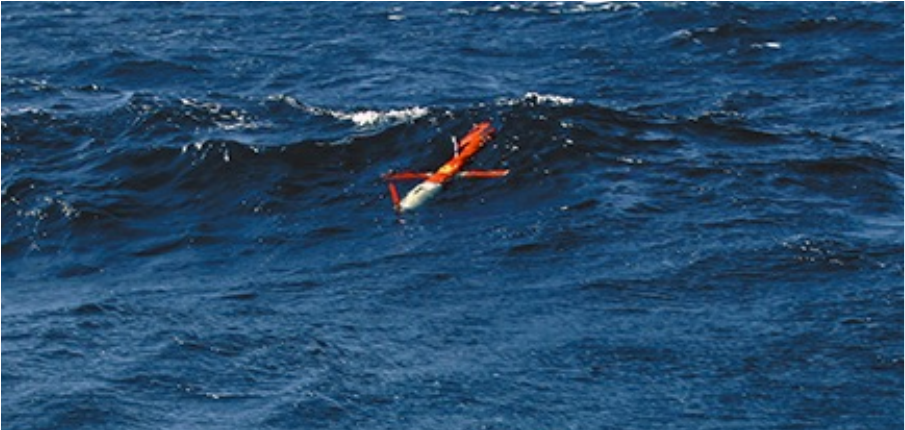}
		\end{minipage}
		\label{fig_glider_11}}
    \subfigure[Underwater glider from NOAA \cite{noaa_glider}]{
    	\begin{minipage}[b]{0.398\textwidth}
   		\includegraphics[width=1\textwidth]{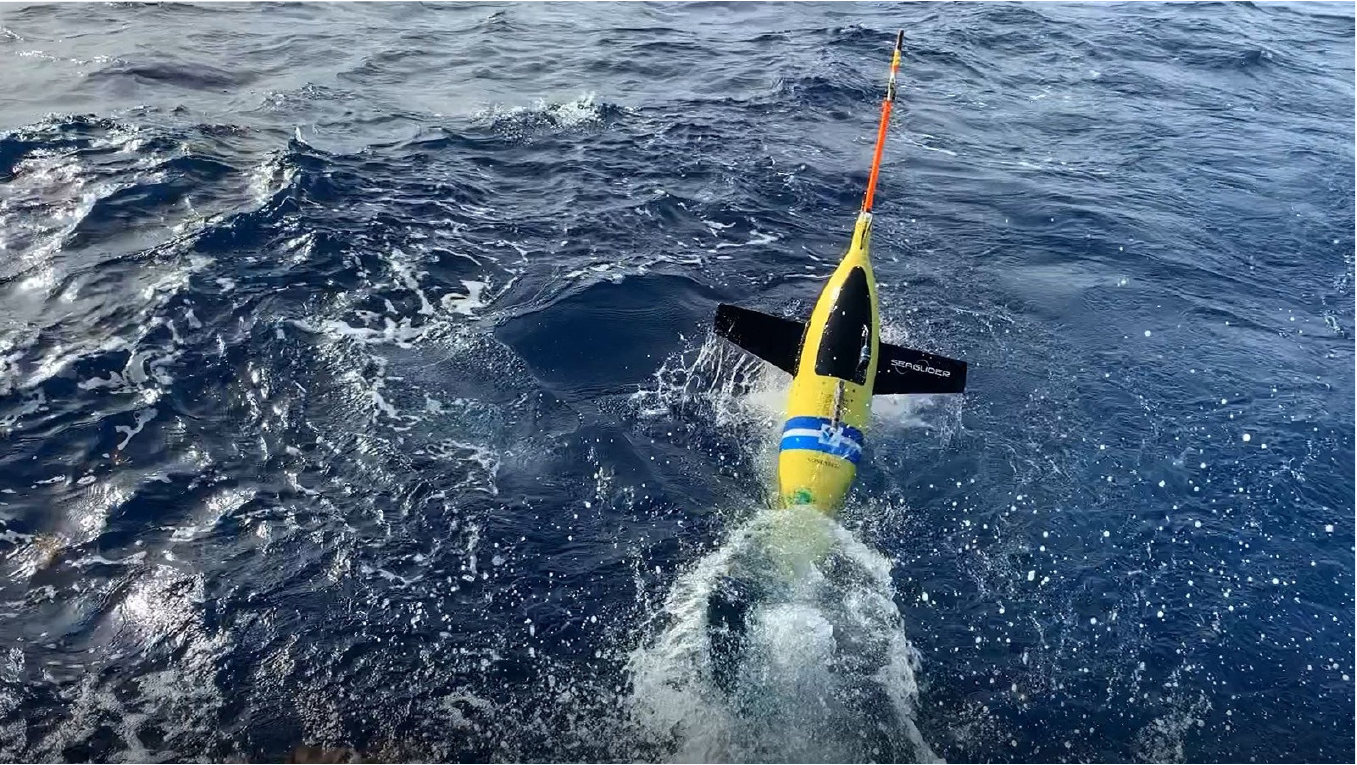}
    	\end{minipage}
		\label{fig_glider_22}}
\caption{Applications of underwater gligers}
\label{fig_glider}
\end{figure}

Practical applications of underwater gliders are shown in\figref{fig_glider}, including the most famous oceanic institution, Woods Hole Oceanographic Institution (WHOI) \cite{whoi_glider}, National Oceanic and Atmospheric Administration (NOAA) \cite{noaa_glider}, and Chinese Academy of Science \cite{cas_glider}.

Unlike traditional AUVs that use propellers or thrusters, underwater gliders change their buoyancy to move up and down in the water column. They reduce their density to ascend and increase it to descend, using small wings to convert vertical motion into forward motion.
Because they rely on buoyancy changes rather than continuous propulsion, underwater gliders are extremely energy-efficient. This allows them to operate for extended periods (weeks to months) and cover long distances (hundreds to thousands of kilometers) on a single battery charge.

The gliders follow a zigzag (sawtooth) path through the water, descending and ascending repeatedly. This enables them to gather data from different depths, making them ideal for profiling the water column. Gliders can be equipped with a variety of sensors and payloads depending on mission objectives, such as temperature, salinity, pressure, dissolved oxygen sensors, acoustic devices, and even cameras.
Gliders are typically pre-programmed to follow specific routes or waypoints. Once deployed, they operate autonomously, occasionally surfacing to transmit collected data to satellites and receive new instructions from operators.
Gliders move slowly, typically around 0.25 to 0.5 meters per second (0.5 to 1 knot). While this makes them unsuitable for fast-response missions, it contributes to their energy efficiency and ability to collect long-term data.

Functions and Applications of Underwater Gliders can be summarized as follows.
\begin{itemize}
 \item Oceanographic Research:
Underwater gliders are widely used in oceanography to collect data on temperature, salinity, chlorophyll concentration, and other parameters. Their ability to cover vast areas and sample at various depths provides valuable information on ocean circulation, water mass characteristics, and climate change impacts.
 \item Environmental Monitoring:
They play a key role in monitoring marine environments, including detecting harmful algal blooms, tracking ocean pollution, and monitoring underwater ecosystems. Their endurance makes them ideal for long-term observations in remote regions, such as the Arctic or deep ocean.
 \item Military and Defense Applications:
Gliders are employed in naval operations for monitoring and surveillance. They can be used to detect submarines, measure underwater acoustics, or track changes in ocean conditions that could impact naval operations.
 \item Data Collection for Climate Studies:
Underwater gliders collect critical data for climate studies, such as temperature profiles of the ocean, which are important for understanding and modeling climate systems, including ocean heat content and its role in global climate change.
 \item Disaster Response:
Gliders can be deployed after natural disasters like hurricanes or oil spills to assess the impact on marine environments and monitor ongoing conditions. Their autonomy allows them to operate in hazardous conditions without putting human lives at risk.
 \item Commercial Applications:
In industries such as oil and gas, gliders are used to monitor underwater infrastructure, map seabed conditions, or provide environmental assessments for offshore installations.
\end{itemize}

With their energy-efficient propulsion system, underwater gliders can perform missions that last weeks to months, far outlasting traditional AUVs or ROVs.
The autonomy and long endurance of gliders make them a cost-effective solution for large-scale oceanographic research and environmental monitoring, reducing the need for expensive manned ships or repeated deployments.
Gliders can travel across entire ocean basins, making them suitable for large-scale surveys. Their ability to sample water at different depths provides comprehensive datasets that are valuable for understanding ocean processes.
Due to their quiet operation and slow movement, gliders are ideal for missions that require stealth, such as naval surveillance or acoustic monitoring.
Gliders can be deployed from ships, small boats, or even from shore, making them highly adaptable to various mission needs and environments.
Underwater gliders are a vital tool for both scientific research and commercial applications, offering unique capabilities for collecting detailed, long-term data across vast and often remote ocean regions.

\subsection{Unmanned surface vehicles (USVs)}

\begin{figure*}[htbp]  
\centering
    \includegraphics[width=9cm]{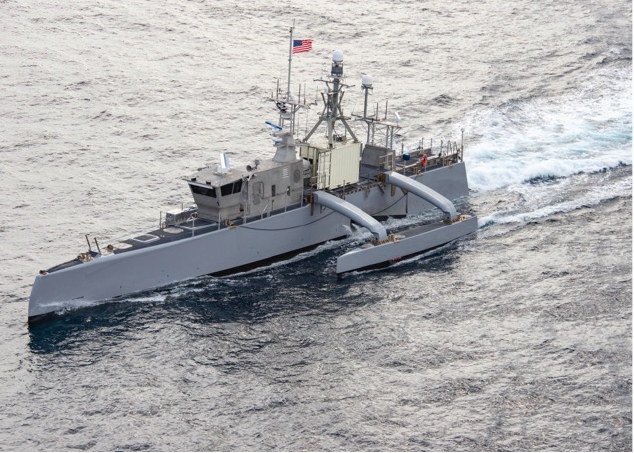}
    \caption{The large USV, "Sea Hunter" \cite{seahunter}}
    \label{fig_large_usv_1}
\end{figure*}

USVs are autonomous or remotely operated vessels that travel on the water's surface. These vehicles are used for various applications, including oceanographic research, military operations, environmental monitoring, and commercial applications (A military USV is shown in \figref{fig_large_usv_1}). USVs offer several advantages, including reduced operational costs, enhanced safety, and the ability to operate in hazardous or remote areas.

\begin{figure}	
\centering
	\subfigure[USV with solar panel \cite{usv_solar1,usv_solar2}]{
	    \begin{minipage}[b]{0.457\textwidth}
	    \includegraphics[width=1\textwidth]{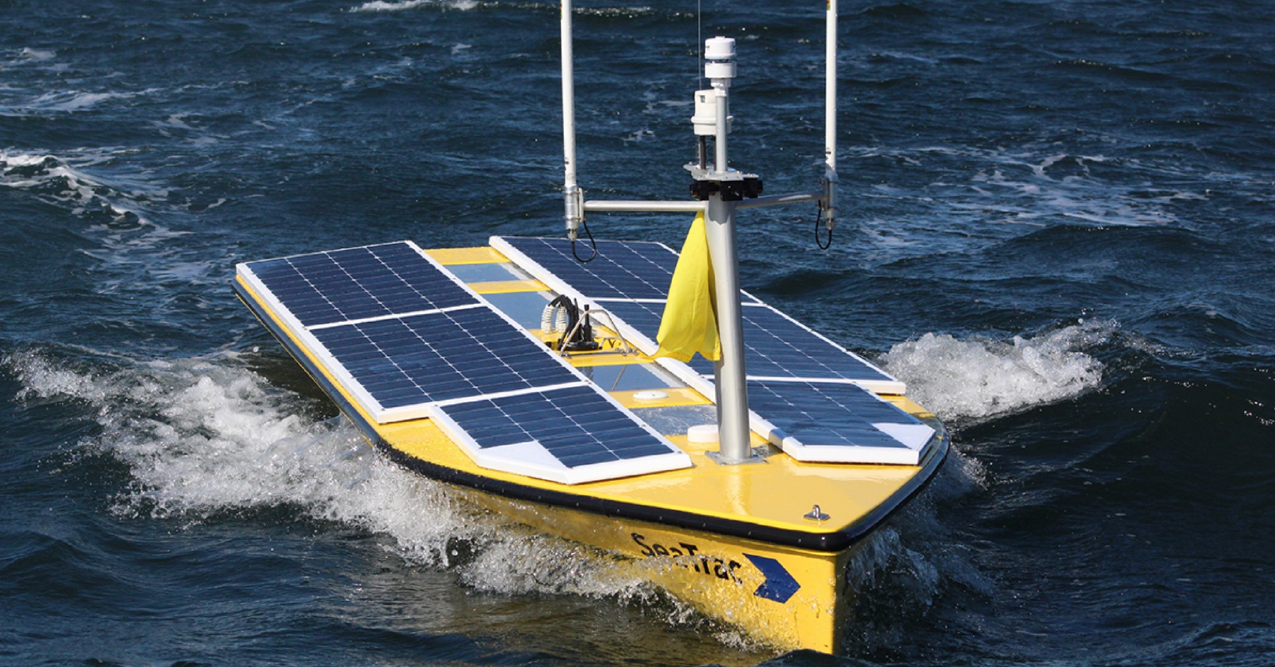}
		\end{minipage}
		\label{fig_usv_4}}
    \subfigure[USV in the shape of catamaran \cite{jiangtun37}]{
    	\begin{minipage}[b]{0.45\textwidth}
   		\includegraphics[width=1\textwidth]{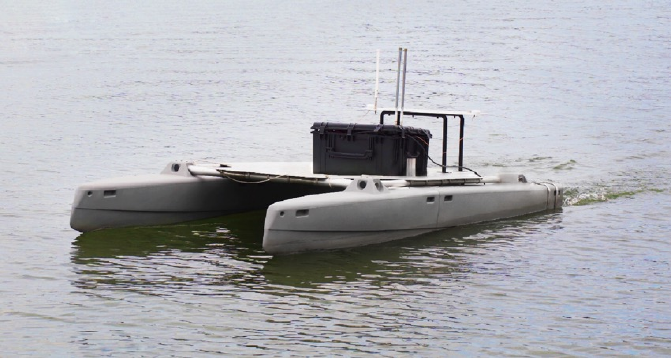}
    	\end{minipage}
		\label{fig_usv_3}}
\caption{name of the whole figure}
\label{fig_usv}
\end{figure}

USVs are typically powered by traditional marine propulsion systems, such as propellers driven by internal combustion engines, electric motors, or hybrid systems. Some advanced USVs use renewable energy sources like solar or wind to extend their operational range and duration(as shown in \figref{fig_usv_4}).
USVs can be fully autonomous, capable of following pre-programmed routes and performing tasks without human intervention, or they can be remotely operated from a control center, allowing for real-time adjustments during missions. Many USVs can switch between these two modes based on mission requirements.
USVs are equipped with GPS, radar, sonar, cameras, and other sensors for precise navigation and obstacle avoidance. Advanced USVs also integrate artificial intelligence (AI) and machine learning algorithms to enhance decision-making during autonomous operations.
USVs rely on satellite communication, radio, and other wireless technologies to transmit data and receive commands. For long-range missions, satellite links ensure continuous communication with operators or data centers.
USVs are designed with modular payload bays, allowing them to carry different types of sensors, cameras, sonar, and other equipment depending on the mission. This flexibility enables a wide range of applications, from scientific research to defense.
USVs come in various hull designs, from traditional mono-hull to catamaran and trimaran configurations (as shown in \figref{fig_usv_3}). These designs are optimized for specific mission requirements, such as speed, stability, and endurance in rough waters.
Depending on the propulsion system, USVs are designed to optimize fuel or energy consumption. Solar-powered USVs, for example, can operate for months at sea, collecting data without requiring refueling or battery replacements.

The main producers of USVs all over the world are listed in the following table (almost all of them are aimed at military application).
\begin{table}[ht]	
\centering
\caption{The main producers of USVs}
	\begin{tabular}{ccc}  
        \toprule
	    Company & Country \\
	    \midrule
	    Kongsberg Maritime &  Norway    \\ 
	    Teledyne Technologies  &  USA   \\ 
	    ECA Group  &  France   \\ 
        Elbit Systems Ltd.  &  USA   \\ 
        Textron Systems  &  USA   \\ 
        ASV Global  &  UK   \\ 
        Atlas Elektronik  &  UK   \\ 
        Rafael Advanced Defense Systems  &  Israel  \\ 
        L3 Harries Technologies  &  USA   \\ 
        Saab  &  Sweden  \\ 
        Fugro  &  Netherlands   \\
        \bottomrule
	\end{tabular}	
\label{table_xxx}
\end{table}

\subsection{Underwater Bionic Robots (UBRs)}

\begin{figure}	
\centering
	\subfigure[An Arowana-like underwater robot \cite{robosea}]{
	    \begin{minipage}[b]{0.45\textwidth}
	    \includegraphics[width=1\textwidth]{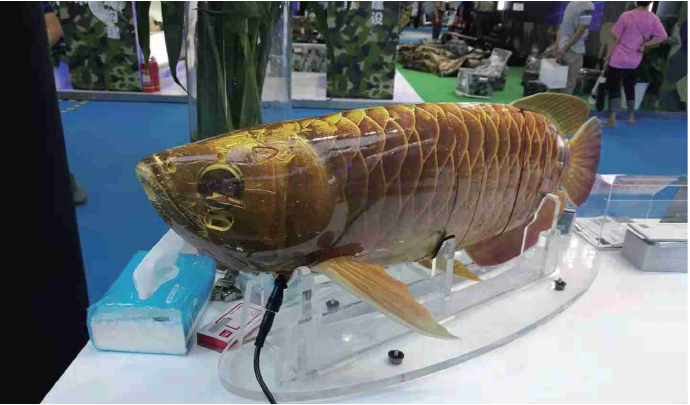}
		\end{minipage}
		\label{fig_arowana_robot}}
    \subfigure[A shark-like underwater robot \cite{robosea}]{
    	\begin{minipage}[b]{0.415\textwidth}
   		\includegraphics[width=1\textwidth]{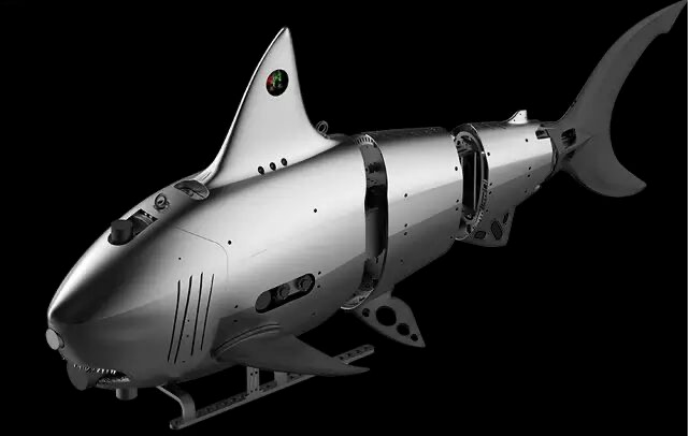}
    	\end{minipage}
		\label{fig_shark_robot}}
\caption{Underwater robots in the shape of fishes}
\label{fig_biorobot}
\end{figure}

\begin{figure}	
\centering
	\subfigure[Underwater robot grasping with a soft hand \cite{softhand}]{
	    \begin{minipage}[b]{0.45\textwidth}
	    \includegraphics[width=1\textwidth]{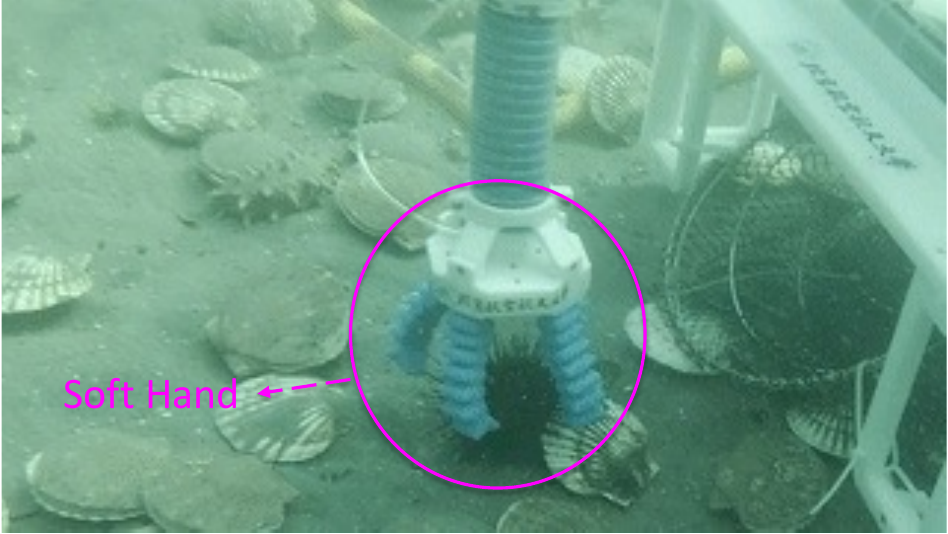}
		\end{minipage}
		\label{fig_soft_hand}}
    \subfigure[Underwater multi-legged robot \cite{crabster}]{
    	\begin{minipage}[b]{0.415\textwidth}
   		\includegraphics[width=1\textwidth]{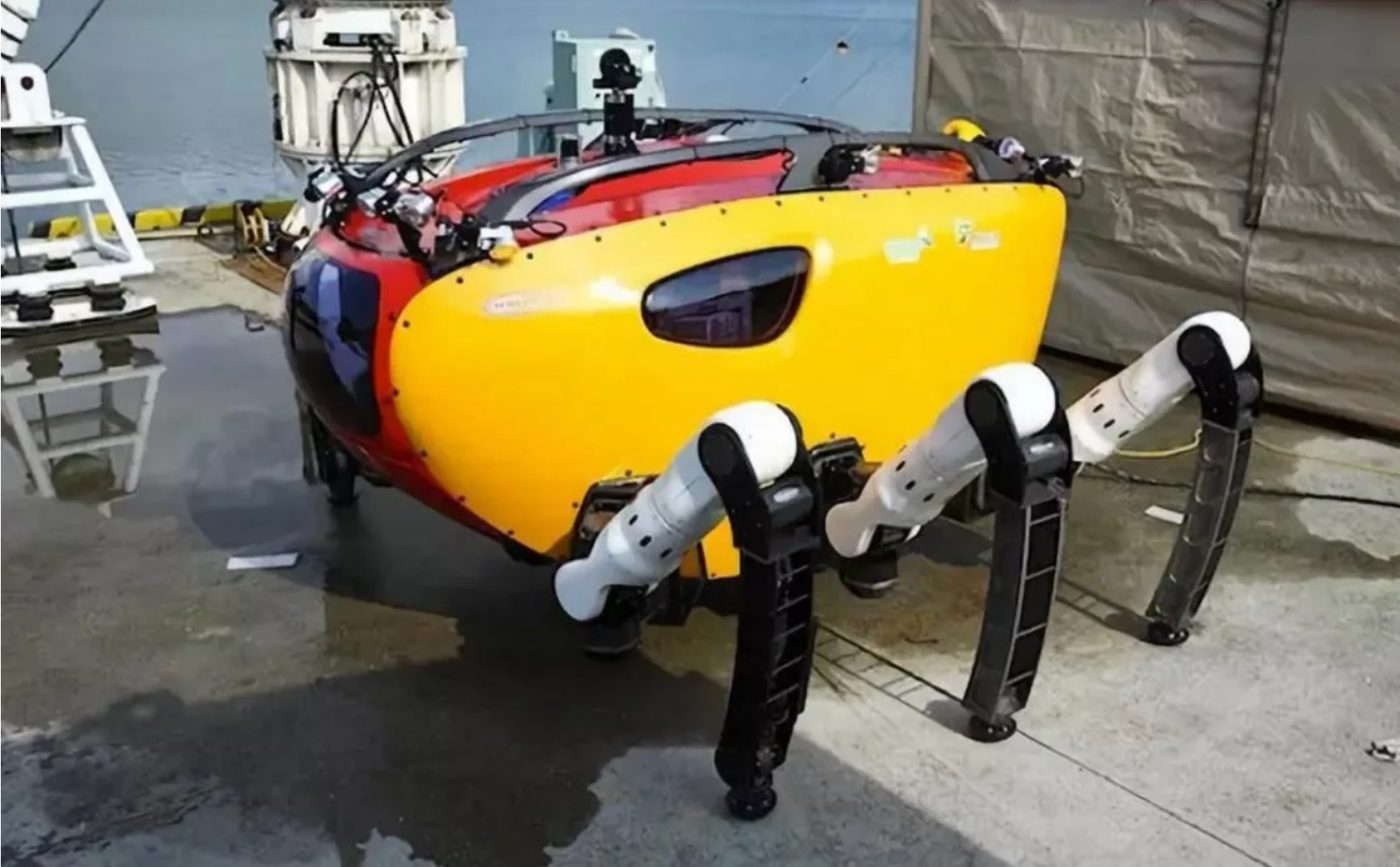}
    	\end{minipage}
		\label{underwater_car_3}}
\caption{Underwater bionic robots}
\label{fig_biorobot2}
\end{figure}

UBRs are designed to emulate the swimming mechanics, agility, and efficiency of fish, octopuses, jellyfish, and other aquatic species, offering unique capabilities for underwater exploration, environmental monitoring, military applications, and research.
 These robots are designed based on the physical characteristics and movement patterns of aquatic creatures. For example, bionic fish robots (as shown in \figref{fig_biorobot} \cite{robosea}) use fin propulsion rather than propellers, mimicking the way real fish swim. Octopus-inspired robots may use soft robotic tentacles for movement and manipulation.
 Many bionic robots use soft materials and flexible structures to replicate the smooth and efficient movements of aquatic animals (as shown in \figref{fig_soft_hand} \cite{softhand} and  \figref{underwater_car_3} \cite{crabster}), enhancing their ability to navigate complex underwater environments without causing damage or noise.
 


By mimicking the efficient movement patterns of marine animals, bionic robots consume significantly less energy compared to traditional AUVs or ROVs with propeller-based propulsion. This allows for longer mission durations and extended operational ranges.
Bionic robots are often designed with streamlined shapes to reduce water resistance (drag), further enhancing their energy efficiency and allowing them to move smoothly through different underwater environments.
Traditional underwater vehicles with thrusters or propellers generate significant noise, which can disturb marine life and reduce stealth in military applications. Bionic robots, by mimicking natural animal movements, produce very little noise, making them ideal for stealth operations or studies of sensitive underwater ecosystems.






\section{Maritime/Underwater Communications}


\subsection{Underwater Wireless Acoustic Communications (UWACs)}

\begin{figure*}[htbp]  
\centering
    \includegraphics[width=12cm]{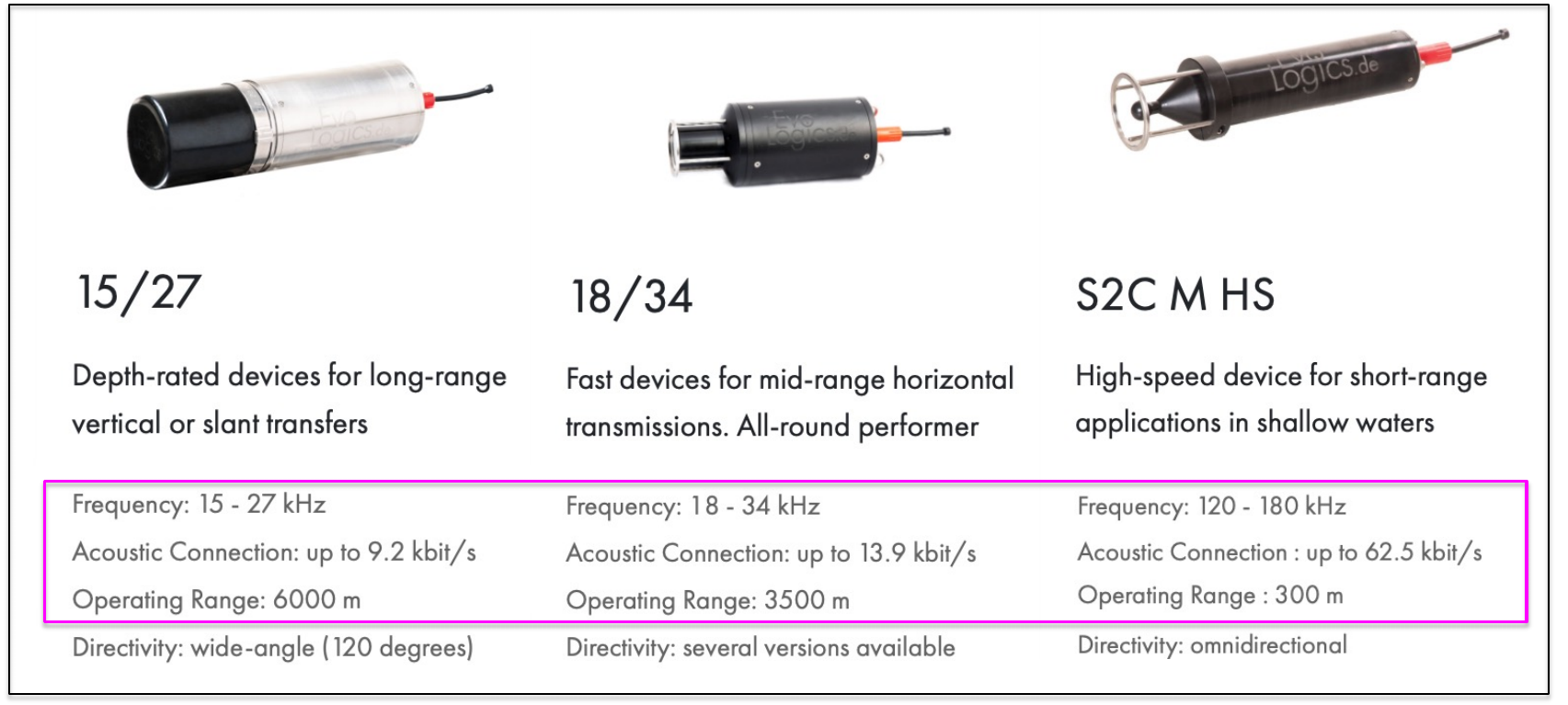}
    \caption{UWAC modems with different frequencies, data rates, and operating ranges}
    \label{fig_uwac_2}
\end{figure*}

As the only way to achieve long-distance ($>$ 100 meters) underwater wireless communication, UWAC is widely used in applications like underwater sensor networks, submarines, marine biology research, and offshore oil exploration. As most of the existing surveys are aimed at the academic research of UWACs in terms of the higher latency, complex and fast-changing channels, serious multipath propagation, etc., here we pay more attention to the practical commercial modems, for example, shown in \figref{fig_uwac_2}, where the communication frequencies, data rates, operating range, and directivity are emphasized.

In the practical application of underwater acoustic communications, choosing the right frequency according to the potential operating range is critical, and it also determines the data rate roughly \tabref{dis-data}. The frequency of operation depends on the intended application, and the required communication range can be classified as:
\begin{itemize}
    \item Low Frequencies (1-10 kHz): Used for long-range communication, up to tens or even hundreds of kilometers. However, these frequencies support low data rates (tens to hundreds of bits per second).
    \item Medium Frequencies (10-100 kHz): Used for medium-range communication (a few kilometers), providing a balance between range and data rate (kilobits per second).
    \item High Frequencies (100 kHz - 1 MHz): Used for short-range communication (tens of meters to a few hundred meters) but provides higher data rates (up to hundreds of kilobits per second or more).
\end{itemize}

\begin{table}
    \centering
       \caption{Description of the underwater communication distance and data rate}
    \begin{tabular}{ccc}
    \toprule
        Description & Distance & Data rate \\
    \midrule
        Super-long distance & $\geq$ 100 km &  $\leq$ 1 kbps\\
        Long distance & 5 $\sim$ 100 km & $\approx$ 10 kbps \\
        Medium distance & 1 $\sim$ 5 km & $\approx$ 20 kbps\\
        Short distance & $\leq$ 1 km  &  $\leq$ 100 kbps \\
    \bottomrule
    \end{tabular}
    \label{dis-data}
\end{table}

\begin{figure*}[htbp]  
\centering
    \includegraphics[width=9cm]{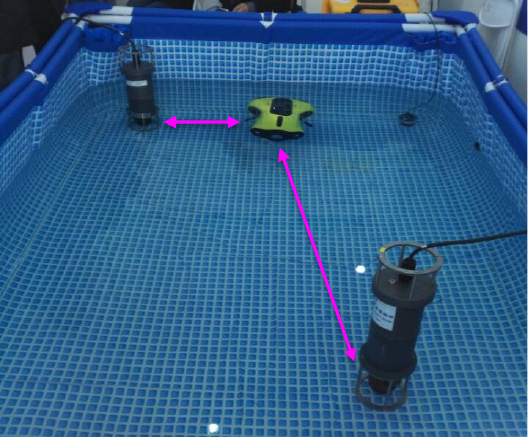}
    \caption{AUV controlling through UWAC}
    \label{fig_uwac_1}
\end{figure*}

A new application for UWAC is shown in \figref{fig_uwac_1}, where an ROV without the tethered cable is connected and controlled by the ground station via UWAC. In this structure, the traditional ROV is transferred to a wireless ROV. One of the more ambitious structures that comes from the demo is an underwater wireless remote-controlled swarm.  Large-scale controllable underwater robot swarms have broad prospects in military and civilian fields, such as large-scale terrain exploration

\subsection{Underwater Wireless Optical Communications (UWOCs)}

\begin{figure}	
\centering
	\subfigure["LUMA X" \cite{hydromea}]{
	    \begin{minipage}[b]{0.4\textwidth}
	    \includegraphics[width=1\textwidth]{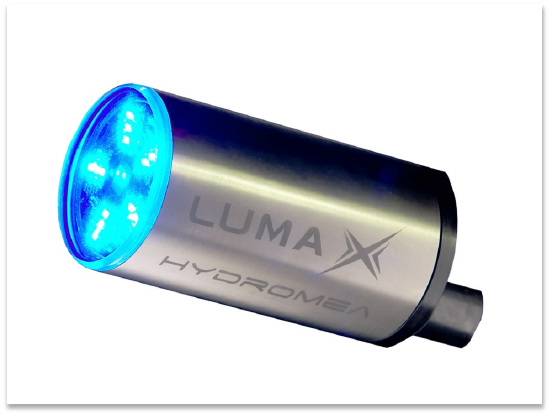}
		\end{minipage}
		\label{fig_luma}}
    \subfigure["BlueComm 100" \cite{sonardyne}]{
    	\begin{minipage}[b]{0.447\textwidth}
   		\includegraphics[width=1\textwidth]{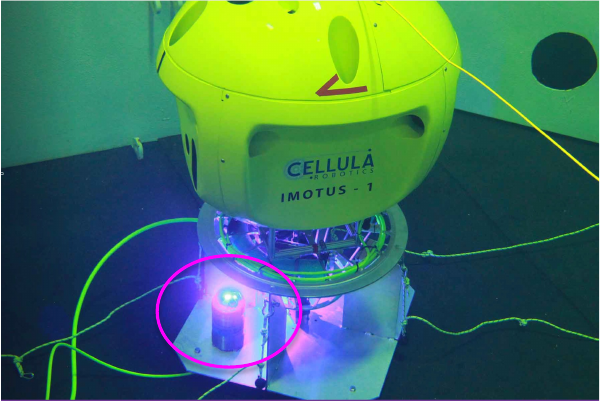}
    	\end{minipage}
		\label{fig_bluecomm}}
\caption{UWOC modems}
\label{fig_uwoc_modems}
\end{figure}

Unlike acoustic or radio frequency (RF) communication systems, which suffer from bandwidth and speed limitations or significant attenuation underwater, underwater optical modems offer high data rates over short to moderate distances. They are used in underwater applications where high-speed communication is necessary, such as real-time data transfer from underwater robots, oceanographic sensors, or subsea infrastructure. The blue-green spectrum (450-550 nm) is optimal for underwater communication, as water absorbs less light in these wavelengths, allowing signals to travel further than other colors of light. Applying the blue-green spectrum, the data rate can be up to 10 Mbps at a distance level of 100 meters. The performance of underwater optical modems is highly dependent on water quality. In clear, clean water, they perform optimally, while in turbid or murky water, performance can degrade significantly due to light scattering and absorption.
Ambient light, particularly sunlight, can introduce noise into optical communication channels. Some systems are optimized for night-time operation or deep-water environments where ambient light is minimal. There are two successful commercial underwater optical communication modems, "LUMA X" and "BlueComm 100," as shown in \figref{fig_uwoc_modems}, which are produced by Hydromea company \cite{hydromea} and Sonardyne \cite{sonardyne}.

\begin{figure*}[htbp]  
\centering
    \includegraphics[width=11cm]{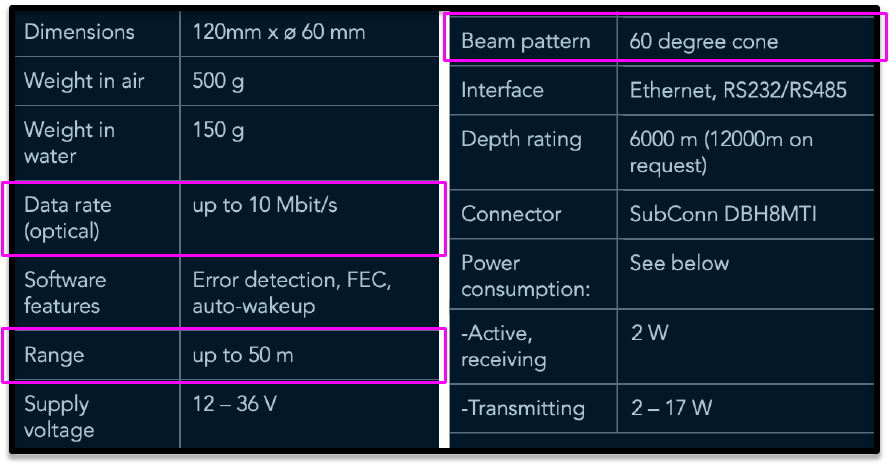}
    \caption{A template of parameters for the UWOC}
    \label{fig_uwoc_2}
\end{figure*}

We can find a template of parameters for the UWOC as shown in \figref{fig_uwoc_2}, where we can see the data rate is quite impressive, and another impressive specific is the wide beam pattern, which can achieve the alignment easily and build the communication link even when the transmitter and the receiver are under unstable conditions like shaking and moving.

\subsection{Underwater Wireless Hybrid Communications (UWHCs)}

Underwater hybrid acoustic and optical modems combine the strengths of both acoustic and optical communication technologies to enable reliable, high-speed, and long-range data transmission underwater. These modems are designed to address the limitations of each technology by offering long-range communication via acoustic signals and high-speed, short-range communication using optical signals, making them highly versatile for various underwater applications. 
In terms of system design, seamless switching between acoustic and optical channels is a key point. Hybrid modems are designed to automatically switch between acoustic and optical modes based on communication needs and environmental conditions. For example, acoustic communication is used for long-distance navigation or control, while optical communication is activated when the modem comes into close proximity with other devices for high-speed data exchange.
Some systems allow for simultaneous communication using both technologies, with acoustic channels handling control commands or status updates and optical channels transmitting high-bandwidth data.

To maximize the range and data rate flexibility, acoustic communication can be designed to provides a range of up to several kilometers but with lower data rates (typically between 100 bps and a few kbps). It is effective in murky or obstructed environments where optical communication might not work.
On the other hand, optical communication can offers very high data rates (from Mbps to Gbps) but is limited to short ranges, typically from a few meters to tens of meters, depending on water clarity. In clear water, the range can be extended up to 100 meters.
At present, there is no such product available on the market. With the support of KAUST, an ambitious project is proposed and attempts to achieve a potential solution for practical application.



\section{Supporting facilities}

Underwater exploration requires a variety of specialized facilities and support systems to ensure effective, safe, and successful operations. These facilities provide the infrastructure, technology, and logistical support needed to explore and study underwater environments, whether for scientific research, resource exploration, environmental monitoring, or defense purposes. In this section, some critical support facilities for underwater exploration is introduced.

\subsection{Submerged Buoys and Surface Buoys }
Submerged buoys and surface buoys are essential tools in marine and underwater environments for monitoring, communication, and data collection. Though both types of buoys serve as platforms for sensors and instrumentation, they have different designs, functions, and applications, depending on whether they are deployed on the water surface or submerged below it.

\begin{figure}	
\centering
	\subfigure[The submerged buoy]{
	    \begin{minipage}[b]{0.46\textwidth}
	    \includegraphics[width=1\textwidth]{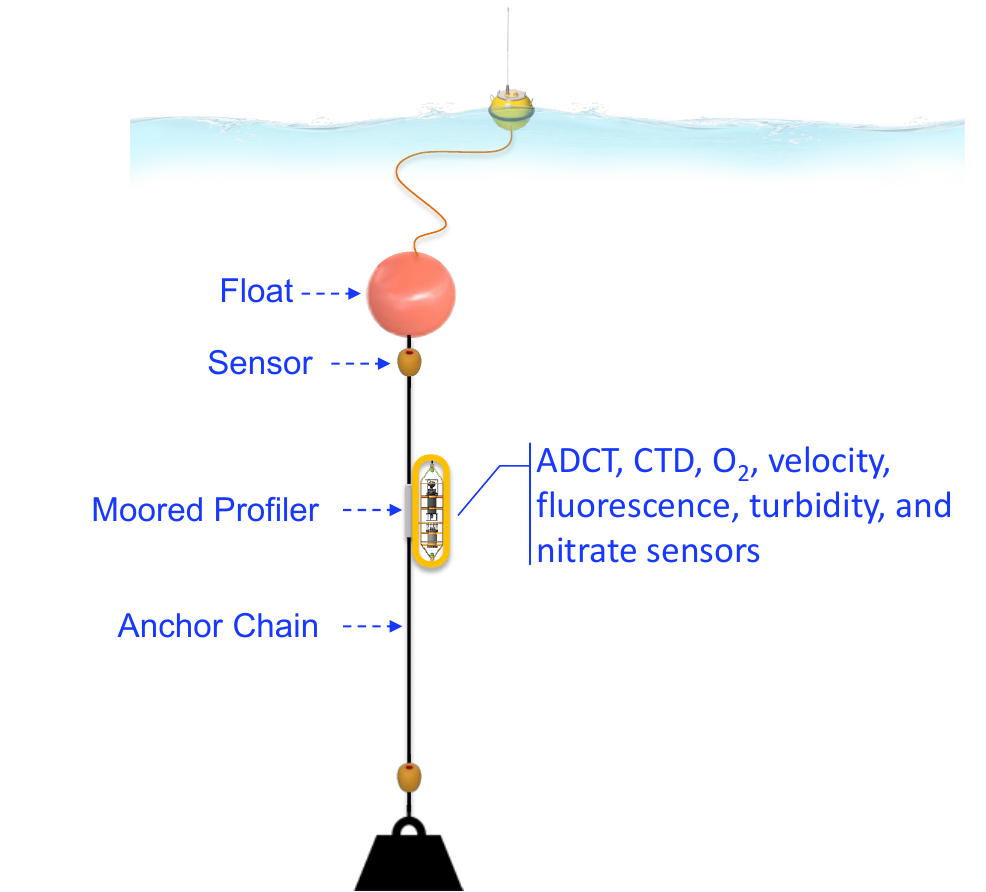}
		\end{minipage}
		\label{fig_submerged_buoy_1}}
    \subfigure[The surface buoy  \cite{buoy3m}]{
    	\begin{minipage}[b]{0.395\textwidth}
   		\includegraphics[width=1\textwidth]{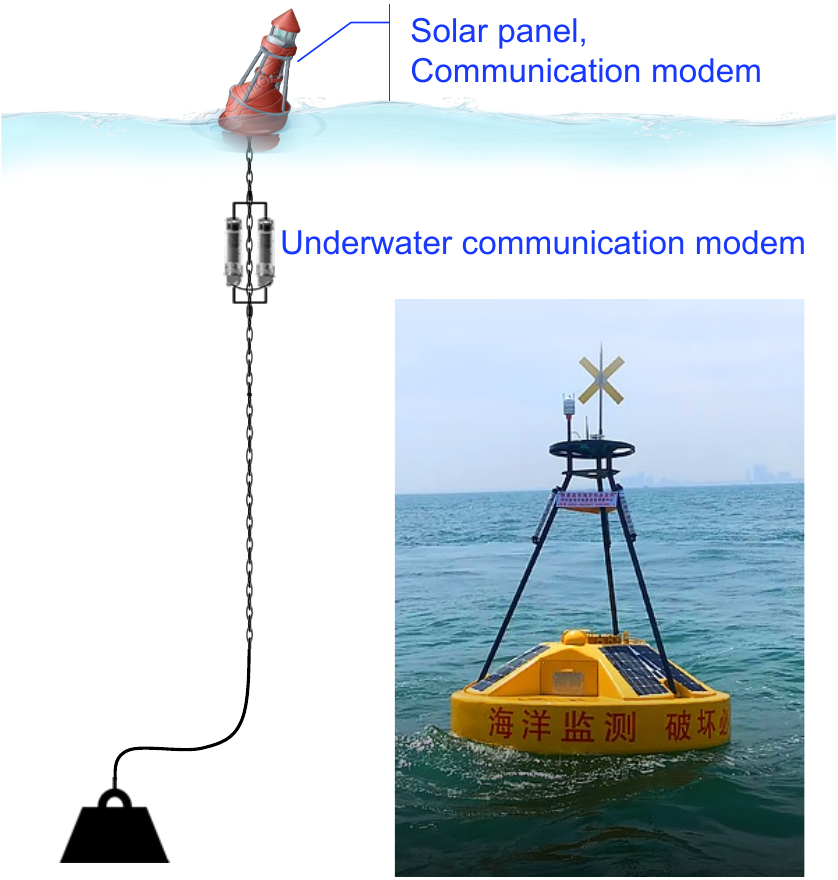}
    	\end{minipage}
		\label{fig_surface_buoy}}
\caption{Different buoys}
\label{fig_buoy}
\end{figure}

Submerged buoys are anchored underwater and remain below the surface (as shown in \figref{fig_submerged_buoy_1}). They are used in scenarios where surface buoys might be impractical due to environmental conditions or where stealth or minimal interference with the surface is required. 
Surface buoys float (as shown in \figref{fig_surface_buoy}) on the water’s surface and serve as platforms for a variety of monitoring, communication, and navigation functions. They are more visible and accessible than submerged buoys, making them suitable for real-time data transmission and broader use in a variety of marine operations.

As the bridge to connect the terrestrial and underwater world, both submerged buoys and surface buoys play vital roles in underwater exploration, oceanographic research, and maritime operations. Submerged buoys are ideal for stealthy, long-term monitoring and data collection below the surface, while surface buoys provide real-time communication, navigation support, and environmental monitoring at the surface level. Their complementary roles ensure a broad range of capabilities for marine research, security, and industrial applications. Comparison of submerged buoys and surface buoys are summarized in \tabref{buoys}.

\begin{table}
    \centering
       \caption{Comparison of submerged buoys and surface buoys}
    \begin{tabular}{m{2.8 cm}<{\centering}|m{5cm}<{\centering}|m{5cm}<{\centering}}
    \toprule
        Aspect & Submerged buoys & Surface buoys \\
    \midrule
        Position & Anchored below the water’s surface &  Floats on the water’s surface\\ \hline
        Visibility & Not visible, used for stealthy or unobtrusive monitoring &  Highly visible with lights, flags, and radar reflectors\\ \hline
        Communication & Acoustic (underwater), can surface for radio or satellite transmission & Radio, satellite, or cellular communication in real time \\ \hline
        Applications & Environmental monitoring, military, tsunami detection, underwater sensor networks & Navigation, climate monitoring, oceanographic research, marine security\\ \hline
        Power supply & Long-lasting batteries, energy harvesting, periodic surfacing  &  Solar panels, rechargeable batteries, fuel cells \\ \hline
        Durability & Solar panels, rechargeable batteries, fuel cells  &  Designed for extreme surface conditions (waves, winds, corrosion) \\ \hline
        Deployment & More complex, requires precise depth control and mooring &  More complex, requires precise depth control and mooring \\ 
    \bottomrule
    \end{tabular}
    \label{buoys}
\end{table}

\subsection{Unmanned deployment and recovery of AUVs}

Since AUV is power by build-in batteries, the energy efficiency is critical for the AUV's performance, especially in some complex application scenarios.

Many underwater missions take place in remote or harsh environments, such as the deep ocean, polar regions, or areas with extreme weather conditions. These areas are often difficult or dangerous for human crews to access, making unmanned operations crucial. Environments like deep-sea exploration, volcanic vents, or areas around oil spills or underwater volcanoes present significant risks to human operators. Unmanned systems allow for safe deployment and recovery of AUVs in such high-risk areas without exposing people to danger.

On the other hand, unmanned systems enable continuous and autonomous deployment and recovery cycles, allowing AUVs to be launched, recovered, recharged, and redeployed without the need for human intervention. This greatly increases the efficiency of long-duration missions, such as environmental monitoring or resource exploration.

\begin{figure*}[htbp]  
\centering
    \includegraphics[width=9cm]{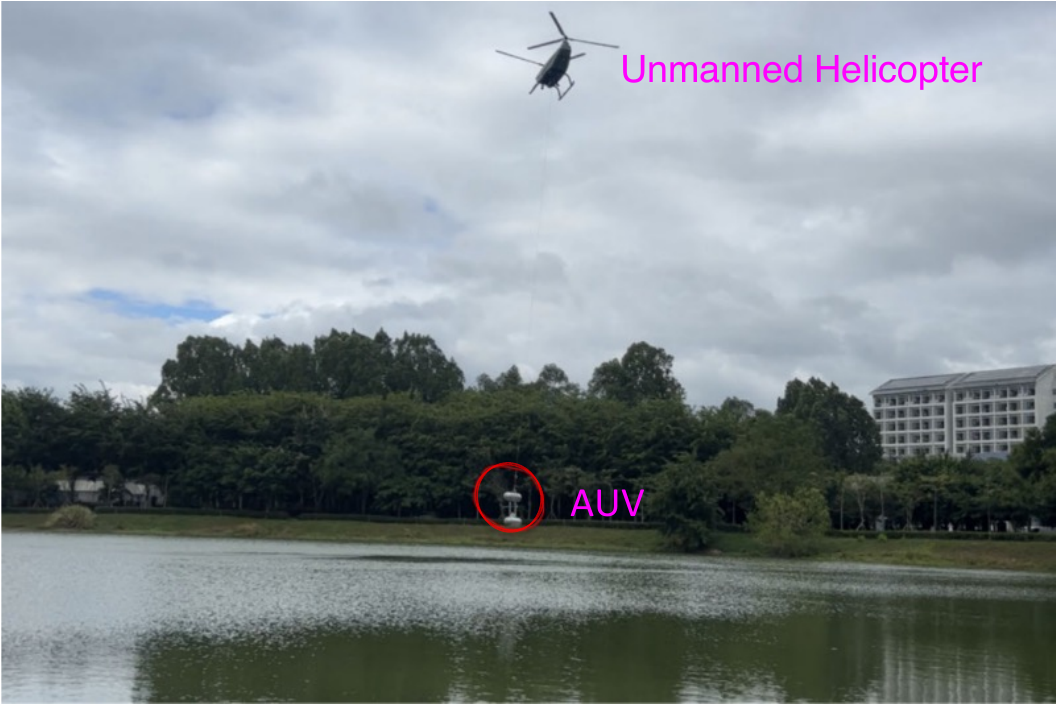}
    \caption{Deploy and recover the AUV with an unmanned helicopter \cite{deployauv}}
    \label{fig_deploy_auv}
\end{figure*}

At the present, USVs, mother ships or autonomous platforms can provided the unmanned deployment and recovery of AUVs. The latest news in \cite{deployauv} shows capacity of unmanned aerial platform to realize unmanned deployment and recovery of AUVs.
The experiment is achieved by Hainan Tropical Ocean University where an unmanned helicopter is used to deploy and recover an AUV \cite{deployauv}, which is shown in \figref{fig_deploy_auv}.

\subsection{Underwater Docking Station (UDS)}

An UDS is a specialized infrastructure that allows AUVs and wireless-ROVs to autonomously dock, recharge, transfer data, and be maintained underwater. Docking stations are designed to extend the operational capabilities of underwater vehicles, enabling long-term missions without the need for surface recovery. These stations are especially valuable in deep-sea, remote, or hazardous environments where frequent surfacing or human intervention is impractical.

Specifics of Underwater Docking Stations can be described as follows.
\begin{itemize}
    \item Design and Structure:
    Underwater docking stations are typically modular, allowing customization based on mission requirements. They consist of docking cradles, power units, data transmission systems, and communication interfaces.
The docking station includes a docking cradle or guidance funnel that enables the AUV to align itself and dock securely. These docking mechanisms are equipped with sensors (e.g., sonar, cameras) to guide the AUV during the docking process.
 Docking stations are constructed using corrosion-resistant and pressure-tolerant materials, such as stainless steel, titanium, or composites, to withstand the harsh underwater environment and high pressures, especially in deep-sea deployments.
 
\item Autonomous Docking Guidance:
The docking station uses various technologies to assist AUVs in locating and docking with: (i)
Sonar and Acoustic Beacons: Acoustic pingers or transponders emit signals that guide the AUV to the docking station. The AUV uses these signals to calculate its position and adjust its approach.
(ii) Cameras and Optical Sensors: In shallow or clear waters, cameras or visual sensors can provide additional guidance to help the AUV align with the docking station.
(iii) Magnetic or Inductive Positioning: In some systems, magnetic or inductive sensors help the AUV lock into the docking station accurately.

\item Power Supply and Recharging:
Docking stations use inductive charging technology, which enables wireless power transfer to recharge AUV batteries without the need for physical connectors. This method is effective in wet and high-pressure environments.
Some docking stations use mechanical connectors for recharging the AUV. The vehicle docks into the station, and electrical contacts deliver power to the AUV’s batteries.
Docking stations are typically connected to a power source, either via underwater cables to a surface ship or shore station, or they may include renewable energy systems (e.g., ocean current turbines or solar panels on surface buoys) to generate power autonomously.

\item Data Transfer and Communication:
While docked, the AUV can transfer mission data (e.g., sonar readings, video footage, sensor data) to the docking station. This data can be relayed to surface vessels or shore-based facilities through: (i) Acoustic Modems, for wireless data transmission while submerged;
(ii) Fiber Optic or Electrical Cables, for direct, high-speed data transfer to the surface.
The docking station can upload new mission instructions or software updates to the AUV while it is docked, enabling continuous, autonomous operations without surfacing.
Docking stations can conduct diagnostics on the AUV’s systems, checking for any malfunctions, battery health, sensor performance, or structural integrity.

\end{itemize}

\begin{figure}	
\centering
	\subfigure[Omnidirectional underwater docking station ]{
	    \begin{minipage}[b]{0.4\textwidth}
	    \includegraphics[width=1\textwidth]{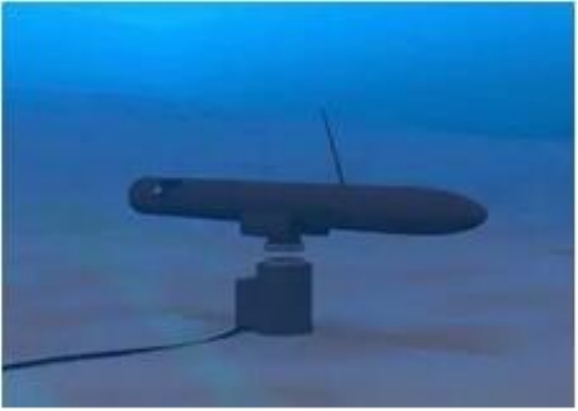}
		\end{minipage}
		\label{fig_underwater_docking_1}}
    \subfigure[Underwater magnetic docking station  \cite{bluelogic}]{
    	\begin{minipage}[b]{0.45\textwidth}
   		\includegraphics[width=1\textwidth]{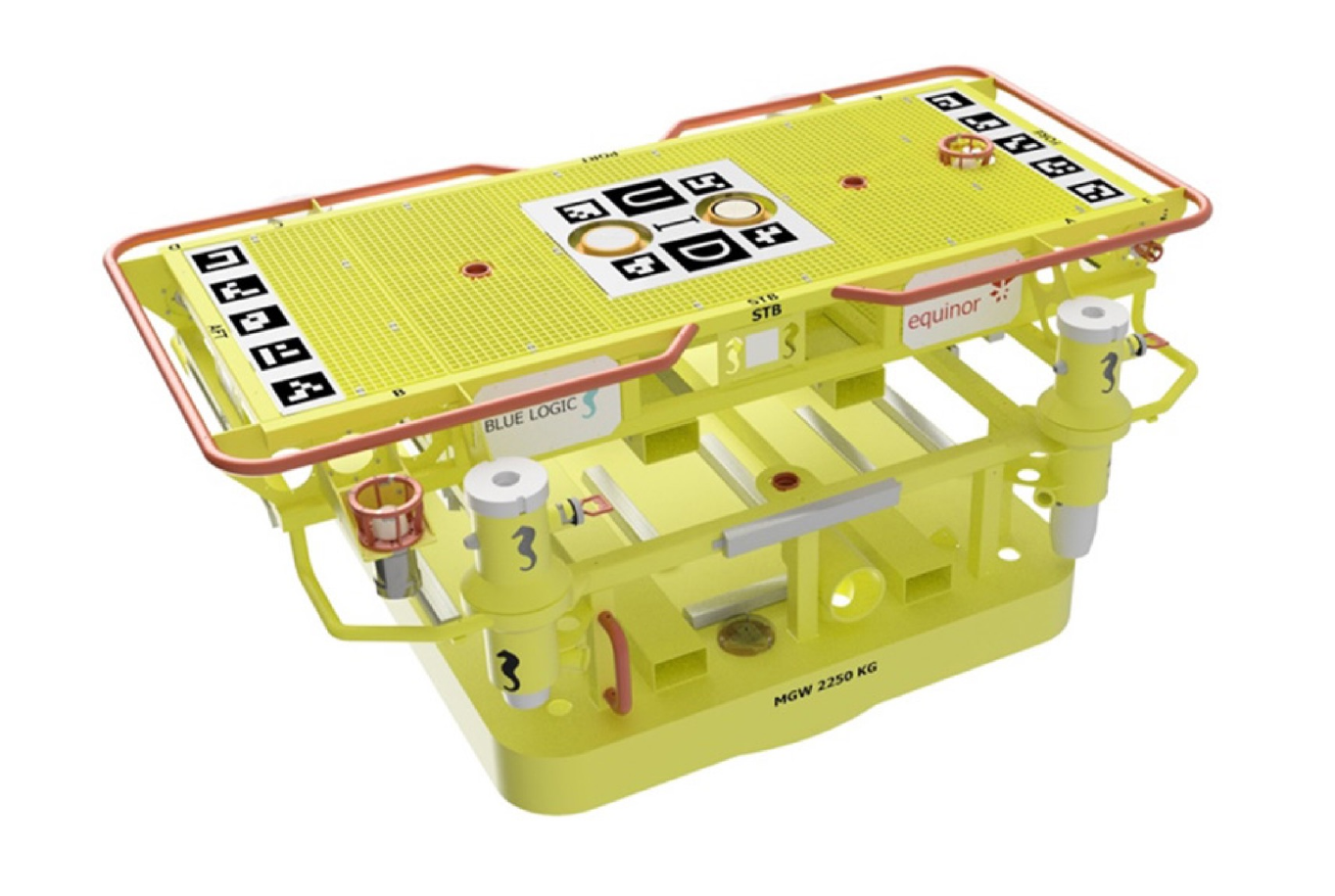}
    	\end{minipage}
		\label{fig_underwater_docking_2}}
\caption{Underwater magnetic docking station}
\label{fig_docking}
\end{figure}

\begin{figure*}[htbp]  
\centering
    \includegraphics[width=11cm]{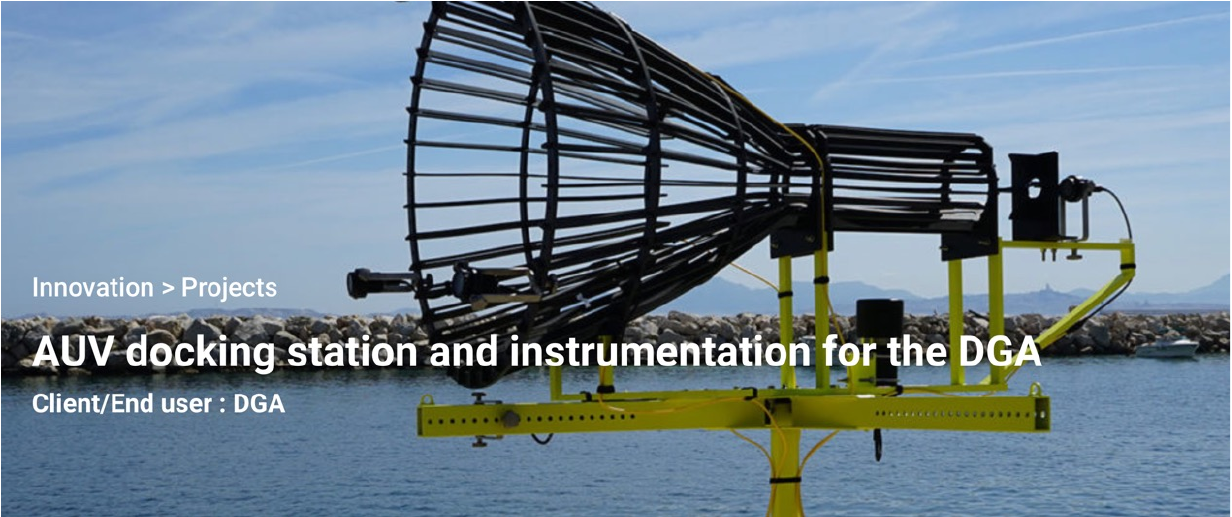}
    \caption{Underwater cage docking station \cite{docking_dga}}
    \label{fig_docking3}
\end{figure*}

At the present, as for as we can see, there is no commercial product of UDS. However, two different structures of UDS's design is presented as shown in \figref{fig_docking} and \figref{fig_docking3}. In  \figref{fig_docking} a docking base is made from magnetic materiel, which assist the station catching the AUV more easily. Furthermore, because magnetism ensures stable contact between the AUV and the base, it means that magnetic wireless charging, short-range optical communication and even wired communication can be achieved. In \figref{fig_docking3}, a cage is designed to carry the AUV. The flared entrance makes it easier for the AUV to enter the base station, even when it is impacted by waves.

\subsection{Underwater Camera and Image Processing Algorithm }

\begin{figure}	
\centering
	\subfigure[Transformation of underwater photo 1]{
	    \begin{minipage}[b]{0.8\textwidth}
	    \includegraphics[width=1\textwidth]{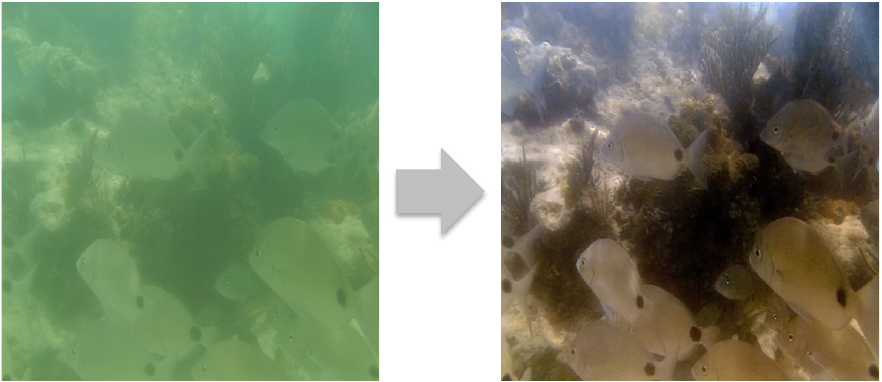}
		\end{minipage}
		\label{fig_camera_2}}
    \subfigure[Transformation of underwater photo 2]{
    	\begin{minipage}[b]{0.8\textwidth}
   		\includegraphics[width=1\textwidth]{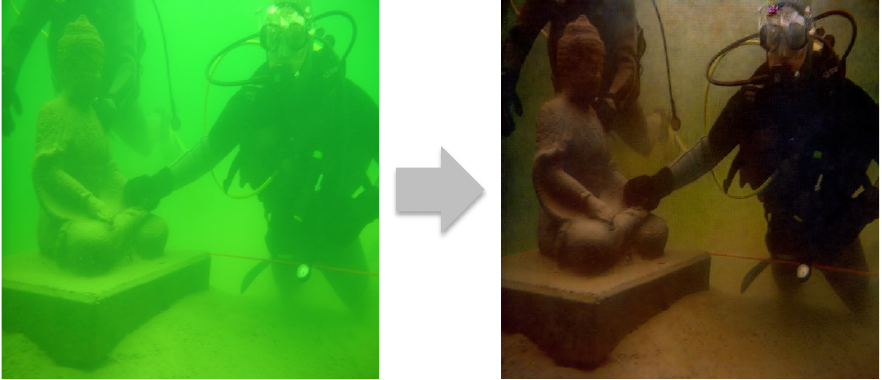}
    	\end{minipage}
		\label{fig_camera_3}}
\caption{Transformation of underwater photos provided by Yanshan University, China}
\label{fig_camera_23}
\end{figure}

Underwater cameras and image processing algorithms have made significant contributions to various fields, including marine biology, underwater exploration, oceanography, environmental monitoring, and industrial applications. These technologies enable the collection, analysis, and interpretation of visual data from underwater environments, where visibility, lighting, and movement pose unique challenges.
Underwater images often suffer from poor visibility due to scattering, absorption, and turbidity in water. Image processing algorithms are used to enhance visibility by reducing haze, restoring color, and improving contrast. A transformation of practical underwater image is shown in \figref{fig_camera_23}.
Algorithms correct distortions caused by the scattering of light underwater, removing haze and blurring to improve image clarity. This is particularly useful in deep-sea environments where visibility is low.
Water absorbs certain wavelengths of light (e.g., red light) more than others, causing color distortion in images. Image processing algorithms correct for this by restoring natural color balance to underwater photographs.



Furthermore, object detection and recognition can be realized by underwater image processing.
Advanced image recognition algorithms are used to automatically identify and classify marine species from underwater footage. This reduces the need for manual labeling and accelerates data analysis in biodiversity studies.
Algorithms are used to detect objects of interest, such as marine debris, shipwrecks, or underwater infrastructure. Automated detection helps in search and recovery missions, as well as in infrastructure inspections.
In industrial applications, image processing algorithms detect anomalies such as cracks, leaks, or corrosion in subsea infrastructure, providing early warnings and reducing the risk of failure.
Algorithms can also stitch together multiple 2D images captured by underwater cameras to create detailed 3D models of underwater environments. These models are used for scientific research, archaeological preservation, and seafloor mapping.
Using stereo cameras and image processing techniques, 3D depth maps of the underwater environment can be generated. These maps are used for navigation, exploration, and understanding complex seafloor topography.






\section{Applied systems}

\subsection{Underwater unmanned vehicle swarm (UUVS)}

UUVSs offer several advantages in various applications, including oceanographic research, environmental monitoring, defense, and underwater infrastructure inspection. Some of the key advantages include:
\begin{itemize}
    \item 
Increased Coverage and Efficiency: A swarm of UUVs can cover a larger area more quickly than a single vehicle, allowing for more comprehensive data collection over vast oceanic regions. This is especially useful for tasks like mapping the seafloor, monitoring marine environments, or searching for underwater objects.

\item Redundancy and Fault Tolerance: In a swarm, if one or a few UUVs fail, the others can continue the mission, making the system more robust. This fault tolerance is particularly important in remote or hazardous environments where retrieval or repair of individual vehicles may be difficult.

\item Improved Data Resolution and Accuracy: UUV swarms can be configured to operate in a coordinated manner, with each vehicle gathering specific data. By combining data from multiple vehicles, a swarm can achieve higher resolution and more accurate measurements, particularly in complex environments like coral reefs, deep-sea trenches, or underwater caves.

\item Scalability and Flexibility: UUV swarms can scale up or down depending on the mission requirements. They can be deployed in various formations or distributed strategies to perform tasks more flexibly. This makes them suitable for both small-scale local missions and large-scale global ocean monitoring.

\item Autonomy and Distributed Intelligence: Swarms of UUVs can operate autonomously, communicating and coordinating with each other to complete tasks more efficiently. Distributed intelligence allows the swarm to adapt to changing environments or mission requirements, such as adjusting their search patterns in real-time based on local conditions.

\item Stealth and Reduced Detection: In defense and surveillance applications, a swarm of smaller UUVs may be harder to detect than a single large vehicle, making them suitable for covert operations or monitoring sensitive areas.

\item Task Specialization: Different UUVs within the swarm can be equipped with various sensors or tools, enabling task specialization. For example, some vehicles might focus on imaging, while others perform water quality analysis or acoustic monitoring. This division of labor enhances the overall capabilities of the swarm.

\end{itemize}

\begin{figure*}[htbp]  
\centering
    \includegraphics[width=11cm]{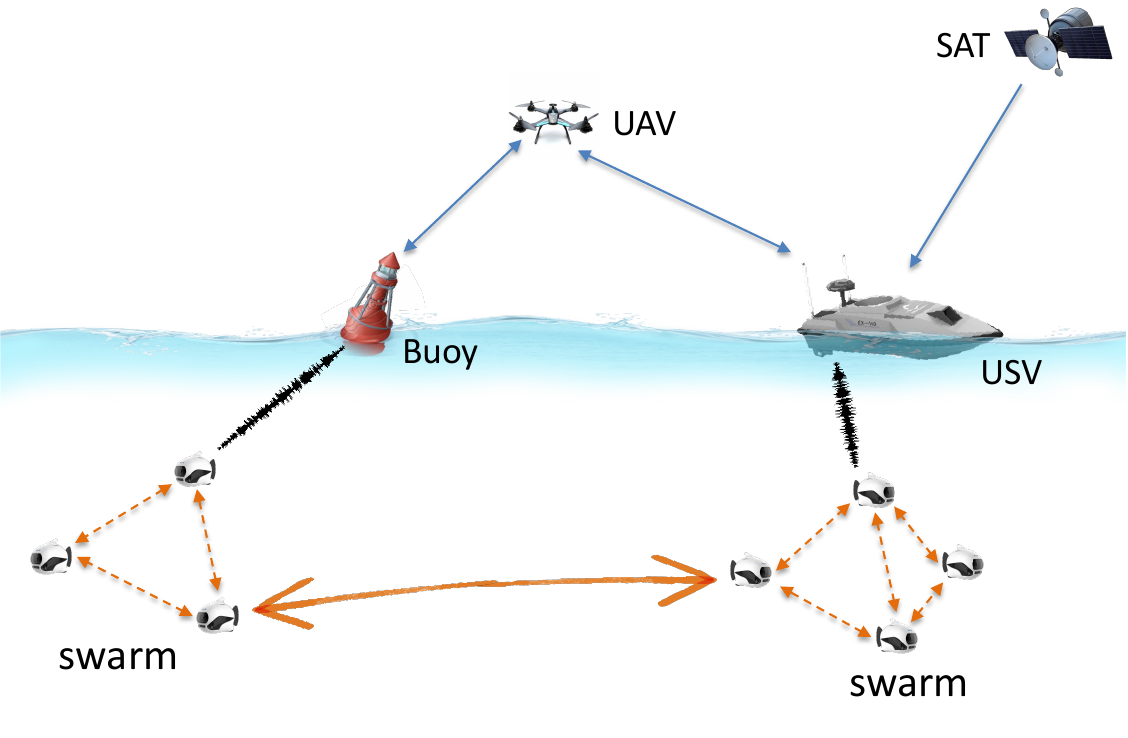}
    \caption{Structure of the underwater swarm}
    \label{fig_underwater_swarm_1}
\end{figure*}

\begin{figure}	
\centering
	\subfigure[An underwater swarm surrounding a boat \cite{underwater_swarm}]{
	    \begin{minipage}[b]{0.38\textwidth}
	    \includegraphics[width=1\textwidth]{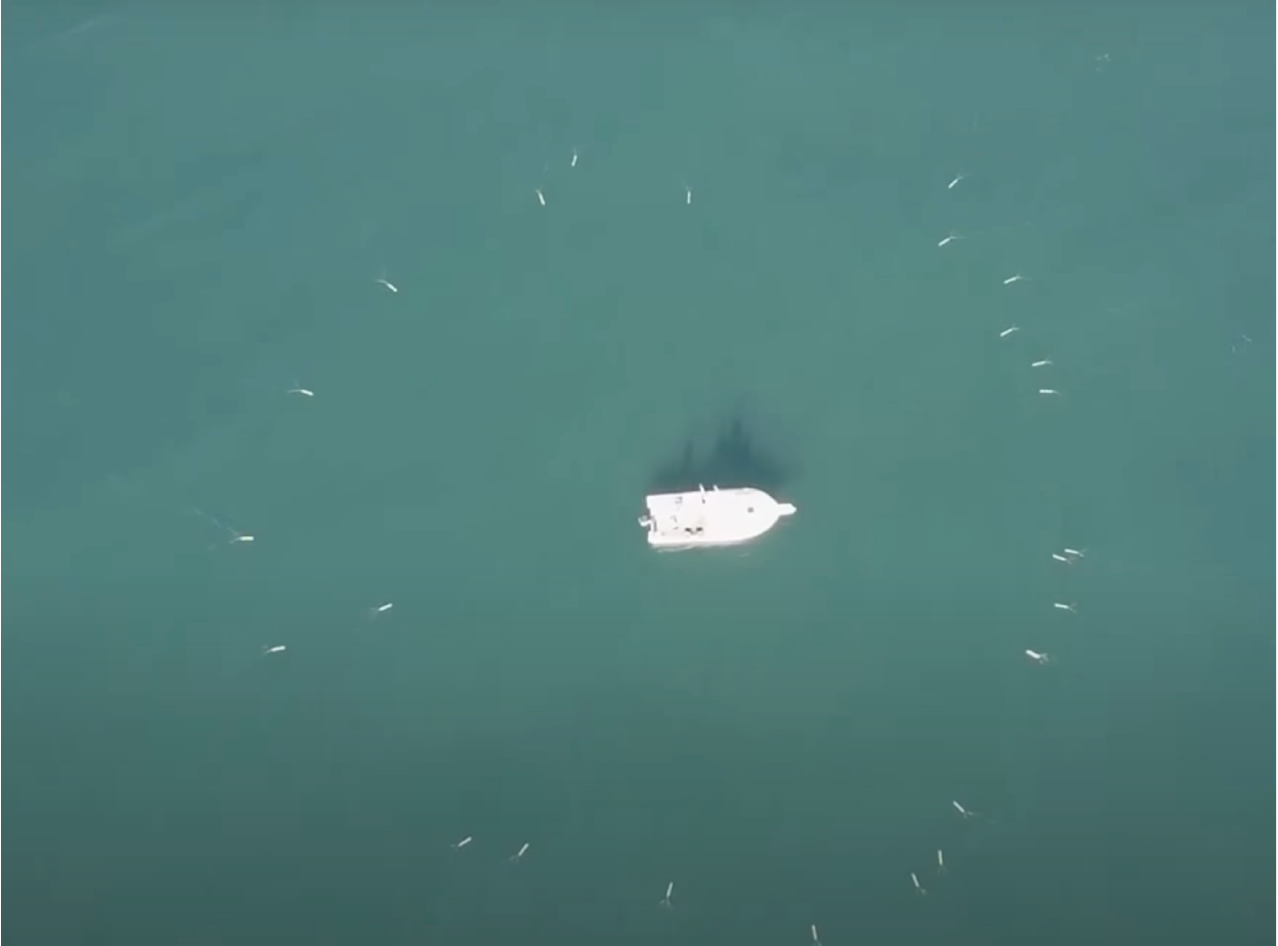}
		\end{minipage}
		\label{fig_underwater_swarm_4}}
    \subfigure[Underwater swarm developed by Harvard University \cite{harvard_swarm}]{
    	\begin{minipage}[b]{0.502\textwidth}
   		\includegraphics[width=1\textwidth]{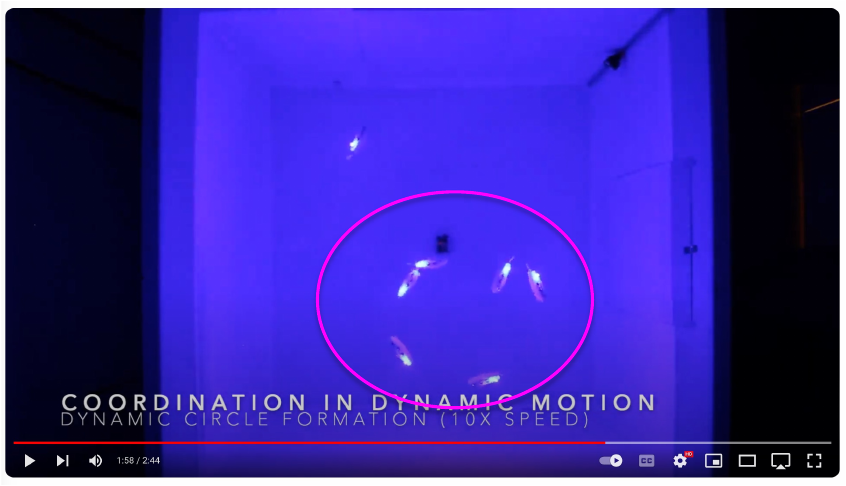}
    	\end{minipage}
		\label{fig_underwater_swarm_2}}
\caption{Experiments for underwater swarms}
\label{fig_underwater_swarm}
\end{figure}

SWARMs project \cite{swarmsproject}  was explored by several countries, including the USA, the UK, etc., which aimed at smart and networking underwater robots in cooperation meshes. Not only AUVs but also ROVs and USVs are applied to further the accessibility and usefulness of the underwater network and make autonomous maritime and offshore operations a viable option for new and existing industries. The structure of UUVS is presented in \figref{fig_underwater_swarm_1}, and there are two experiments (as shown in\figref{fig_underwater_swarm}) achieved by Aquabotix \cite{underwater_swarm}, an Australian company, and Harvard University \cite{harvard_swarm}. Regarding the available product of underwater swarms, as we know, the only commercial model of  UUVS, "VERTEX", is produced by Hydromea company \cite{hydromeaswarm2,hydromeaswarm}.

\subsection{Underwater mapping and detection systems}

\begin{figure*}[htbp]  
\centering
    \includegraphics[width=14cm]{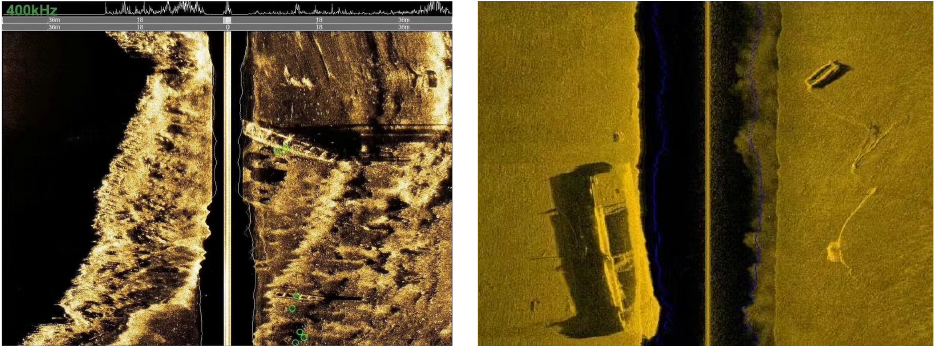}
    \caption{Underwater mapping and detection\cite{zhonghaida}}
    \label{fig_underwater_mapping_detection_12}
\end{figure*}

The development of underwater mapping and detection systems is motivated by a variety of scientific, environmental, commercial, and defense needs. These systems enable the exploration, monitoring, and utilization of the underwater environment in ways that were previously impossible or inefficient. Here are the primary motivations driving the advancement of underwater mapping and detection systems. The motivation can be described as:
\begin{itemize}
    \item Mapping the Seafloor: A large portion of Earth's seafloor remains unexplored. Developing advanced underwater mapping systems helps scientists create detailed maps of the ocean floor, discovering underwater features like ridges, trenches, and volcanoes \cite{blueprint}.
\item Marine Biology: Mapping and detecting underwater habitats allow for better understanding and conservation of marine ecosystems, including coral reefs, fish populations, and biodiversity \cite{esri}.
\item Geological Studies: Underwater detection systems help identify geological formations, monitor tectonic activity, and study underwater mineral deposits and hydrothermal vents, contributing to our understanding of Earth's geology.
\item Improving Navigational Charts: Accurate underwater mapping is essential for creating and updating navigational charts, especially for coastal regions, shipping lanes, and shallow waters where there is a high risk of grounding \cite{3dat}.
\item Hazard Detection: Underwater detection systems help identify obstacles such as shipwrecks, submerged rocks, or other hazards to navigation, ensuring the safety of ships and submarines. A practical template for hazard Detection is shown in \figref{fig_underwater_mapping_detection_12}.
\item Port and Harbor Management: Ports and harbors require regular mapping to monitor sedimentation, underwater infrastructure, and ensure safe passage for vessels\cite{coda}.
\end{itemize}




\subsection{Wearable underwater localization and communication systems (WULCS)}

\begin{figure*}[htbp]  
\centering
    \includegraphics[width=13cm]{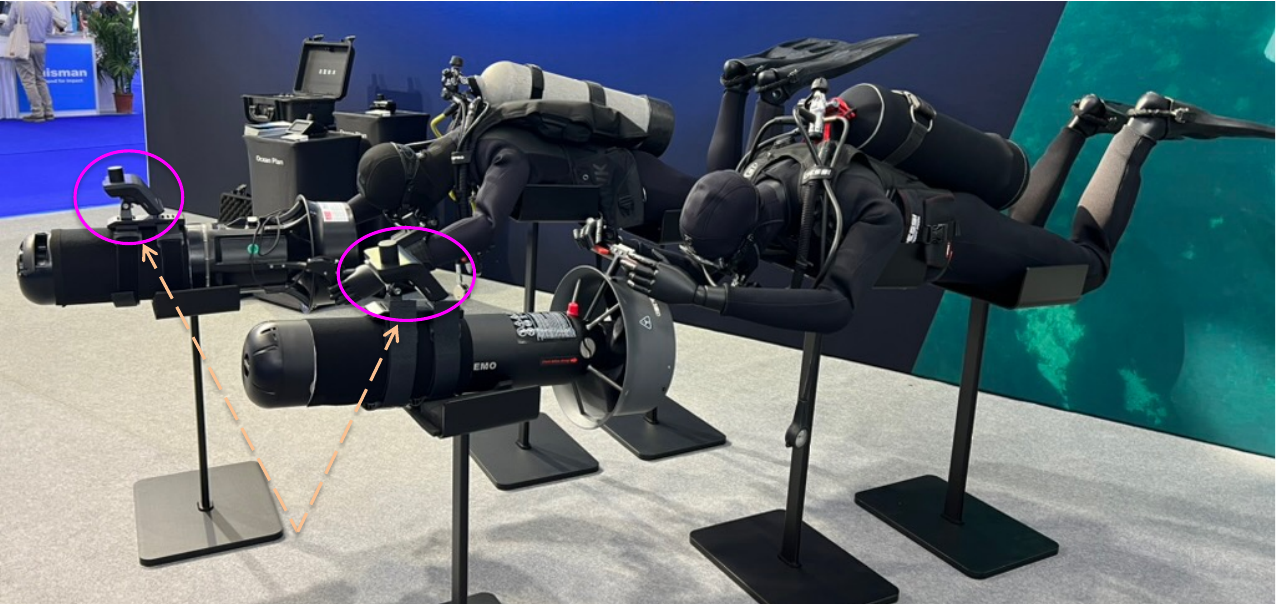}
    \caption{The wearable localization communication device \cite{wearable_localization}}
    \label{fig_wearable_localization}
\end{figure*}

Wearable underwater localization and communication systems are advanced technologies designed to enable real-time tracking, communication, and data transmission for divers, underwater researchers, military personnel, or marine professionals working in underwater environments. These systems provide reliable positioning and communication in challenging underwater conditions where traditional methods, such as GPS and radio communication, are ineffective due to the properties of water. A practical WULCS is shown in \figref{fig_wearable_localization}, where the acoustic signal is applied for underwater positioning and communication. According to the application scenario and the requirement of the mission, the common wearable localization techniques include:
\begin{itemize}
    \item Ultra-Short Baseline (USBL): A surface station (on a boat or buoy) sends acoustic signals to the wearable device, which responds. By measuring the signal's travel time and direction, the surface station can calculate the wearer's position.
\item Long Baseline (LBL): Multiple fixed acoustic beacons are deployed on the seafloor. The wearable system communicates with these beacons, and the position is triangulated based on the time delays of the signals from different beacons.
\item  Short Baseline (SBL): Similar to USBL, but with shorter distances and typically used for small areas like near platforms or underwater structures. 
\end{itemize}

Besides the localization, communication systems can be achieved in terms of text and data transmission, and multi-user interaction. Furthermore, some wearable systems are equipped with HMDs that allow divers to visualize location data, messages, and navigation information in real time. This enables hands-free navigation and communication.
 Wearable systems often include wrist or arm-mounted displays to show vital information such as depth, location, direction, and incoming messages.
In low-visibility conditions, wearable systems may include haptic feedback mechanisms (e.g., vibrations) to provide directional cues or alert divers to critical messages without the need for visual displays.

\section{Conclusion}

The advancements in underwater vehicles and enabling technologies are revolutionizing our ability to explore, monitor, and sustainably manage the world’s oceans. This survey provides a comprehensive classification and analysis of state-of-the-art underwater vehicles, including Autonomous Underwater Vehicles (AUVs), Remotely Operated Vehicles (ROVs), Hybrid Vehicles, and wearable systems, as well as the communication, navigation, and sensing technologies that drive their operations.

Key trends indicate a growing shift toward greater autonomy, extended mission duration, improved data collection, and real-time communication capabilities. Autonomous vehicles are becoming more capable of performing complex tasks, from deep-sea exploration to environmental monitoring, while hybrid systems integrate the strengths of multiple vehicle types to meet the diverse needs of ocean industries. The integration of AI, machine learning, and advanced sensor technologies is enabling smarter, more efficient vehicles capable of processing vast amounts of data, enhancing their decision-making and adaptability in unpredictable underwater environments.

As we move forward, the concept of the Smart Ocean—an interconnected ecosystem of underwater vehicles, sensors, and communication networks—promises to play a pivotal role in addressing critical global challenges. From resource management and environmental protection to climate change monitoring and maritime security, these technologies are essential for ensuring the sustainable use of the ocean’s resources. Future developments will focus on improving energy efficiency, reducing operational costs, and enhancing the scalability of underwater systems to meet the demands of both scientific research and industrial applications.

In conclusion, the continuous innovation in underwater vehicle technologies and the growing application of smart systems are propelling us toward a more connected, informed, and sustainable ocean future.



\bibliographystyle{IEEEtran}
\bibliography{reference}   

\end{document}